\documentclass[twocolumn]{aastex631}

\DeclareUnicodeCharacter{2212}{-}
\newcommand{\sigjit}{\ensuremath{\sigma_{\mathrm{jitter}}}}
\def\MgMedian{$15.0 \pm 1.3\, M_\oplus$}
\def\MhMedian{$203 \pm 16\, M_\oplus$}

\begin{document}

\title{Updated Masses for the Gas Giants in the Eight-Planet Kepler-90 System Via Transit-Timing Variation and Radial Velocity Observations}

\author[0009-0002-1850-2204]{David E. Shaw}
\affil{Department of Physics and Astronomy, 225 Nieuwland Science Hall, University of Notre Dame, Notre Dame, IN 46556, USA}

\author[0000-0002-3725-3058]{Lauren M. Weiss}
\affil{Department of Physics and Astronomy, 225 Nieuwland Science Hall, University of Notre Dame, Notre Dame, IN 46556, USA}

\author[0000-0002-0802-9145]{Eric Agol}
\affil{Department of Astronomy, University of Washington, Seattle, WA 98195, USA}

\author[0000-0001-6588-9574]{Karen A.\ Collins}
\affiliation{Center for Astrophysics \textbar \ Harvard \& Smithsonian, 60 Garden Street, Cambridge, MA 02138, USA}

\author[0000-0003-1464-9276]{Khalid Barkaoui}
\affiliation{Astrobiology Research Unit, Universit\'e de Li\`ege, 19C All\'ee du 6 Ao\^ut, 4000 Li\`ege, Belgium}
\affiliation{Department of Earth, Atmospheric and Planetary Science, Massachusetts Institute of Technology, 77 Massachusetts Avenue, Cambridge, MA 02139, USA}
\affiliation{Instituto de Astrof\'isica de Canarias (IAC), Calle V\'ia L\'actea s/n, 38200, La Laguna, Tenerife, Spain}

\author[0000-0001-8621-6731]{Cristilyn N.\ Watkins}
\affiliation{Center for Astrophysics \textbar \ Harvard \& Smithsonian, 60 Garden Street, Cambridge, MA 02138, USA}

\author[0000-0001-8227-1020]{Richard P. Schwarz}
\affiliation{Center for Astrophysics \textbar \ Harvard \& Smithsonian, 60 Garden Street, Cambridge, MA 02138, USA}

\author[0009-0009-5132-9520]{Howard M. Relles}
\affiliation{Center for Astrophysics \textbar \ Harvard \& Smithsonian, 60 Garden Street, Cambridge, MA 02138, USA}

\author[0000-0003-2163-1437]{Chris Stockdale}
\affiliation{Hazelwood Observatory, Australia}

\author[0000-0003-0497-2651]{John F.\ Kielkopf}
\affiliation{Department of Physics and Astronomy, University of Louisville, Louisville, KY 40292, USA}

\author{Fabian Rodriguez Frustaglia}
\affiliation{Frustaglia Private Observatory, Spain}

\author[0000-0001-6637-5401]{Allyson Bieryla}
\affiliation{Center for Astrophysics \textbar \ Harvard \& Smithsonian, 60 Garden Street, Cambridge, MA 02138, USA}

\author[0000-0002-0145-5248]{Joao Gregorio}
\affiliation{Crow Observatory, Portalegre, Portugal}

\author[0009-0005-5648-7107]{Owen Mitchem}
\affiliation{Pine Mountain Observatory, Institute for Fundamental Science, Department of Physics, University of Oregon, Eugene, OR 97403, USA}

\author[0009-0009-7566-8420]{Katherine Linnenkohl}
\affiliation{Pine Mountain Observatory, Institute for Fundamental Science, Department of Physics, University of Oregon, Eugene, OR 97403, USA}

\author[0000-0003-3184-5228]{Adam Popowicz}
\affiliation{Silesian University of Technology, Department of Electronics, Electrical Engineering and Microelectronics, Akademicka 16, 44-100 Gliwice, Poland}

\author[0000-0001-8511-2981]{Norio Narita}
\affiliation{Komaba Institute for Science, The University of Tokyo, 3-8-1 Komaba, Meguro, Tokyo 153-8902, Japan}
\affiliation{Astrobiology Center, 2-21-1 Osawa, Mitaka, Tokyo 181-8588, Japan}
\affiliation{Instituto de Astrofisica de Canarias (IAC), 38205 La Laguna, Tenerife, Spain}

\author[0000-0002-4909-5763]{Akihiko Fukui}
\affiliation{Komaba Institute for Science, The University of Tokyo, 3-8-1 Komaba, Meguro, Tokyo 153-8902, Japan}
\affiliation{Instituto de Astrofisica de Canarias (IAC), 38205 La Laguna, Tenerife, Spain}

\author[0000-0003-1462-7739]{Michaël Gillon}
\affiliation{Astrobiology Research Unit, Université de Liège, 19C Allée du 6 Août, 4000 Liège, Belgium}
  
\author[0000-0003-3904-6754]{Ramotholo Sefako} 
\affiliation{South African Astronomical Observatory, P.O. Box 9, Observatory, Cape Town 7935, South Africa}

\author[0000-0002-1836-3120]{Avi Shporer}
\affiliation{Department of Physics and Kavli Institute for Astrophysics and Space Research, Massachusetts Institute of Technology, Cambridge, MA 02139, USA}

\author{Adam Lark}
\affiliation{Hamilton College, 198 College Hill Rd, Clinton, NY 13413, USA}

\author{Amelie Heying}
\affiliation{Hamilton College, 198 College Hill Rd, Clinton, NY 13413, USA}

\author{Isa Khan}
\affiliation{Hamilton College, 198 College Hill Rd, Clinton, NY 13413, USA}

\author{Beibei Chen}
\affiliation{Hamilton College, 198 College Hill Rd, Clinton, NY 13413, USA}

\author[0009-0008-9182-7471]{Kylee Carden}
\affiliation{Department of Astronomy, The Ohio State University, 140 W 18th Ave, Columbus, OH 43210, USA}

\author{Donald M. Terndrup}
\affiliation{Department of Astronomy, The Ohio State University, 140 W 18th Ave, Columbus, OH 43210, USA}

\author{Robert Taylor}
\affiliation{Department of Astronomy, The Ohio State University, 140 W 18th Ave, Columbus, OH 43210, USA}

\author{Dasha Crocker}
\affiliation{Department of Astronomy, The Ohio State University, 140 W 18th Ave, Columbus, OH 43210, USA}

\author[0000-0002-3247-5081]{Sarah Ballard}
\affiliation{Department of Astronomy, University of Florida, Gainesville, FL 32611, USA}

\author[0000-0003-3750-0183]{Daniel C. Fabrycky}
\affiliation{Department of Astronomy and Astrophysics, University of Chicago, Chicago, IL 60637, USA}

\begin{abstract}
The eight-planet Kepler-90 system exhibits the greatest multiplicity of planets found to date. All eight planets are transiting and were discovered in photometry from the NASA Kepler primary mission. The two outermost planets, g ($P_g = 211 \, \textrm{d}$) and h  ($P_h = 332 \, \textrm{d}$) exhibit significant transit-timing variations (TTVs), but were only observed 6 and 3 times respectively by Kepler. These TTVs allow for the determination of planetary masses through dynamical modeling of the pair's gravitational interactions, but the paucity of transits allows a broad range of solutions for the masses and orbital ephemerides. To determine accurate masses and orbital parameters for planets g and h, we combined 34 radial velocities (RVs) of Kepler-90, collected over a decade, with the Kepler transit data. We jointly modeled the transit times of the outer two planets and the RV time series, then used our two-planet model to predict their future times of transit. These predictions led us to recover a transit of Kepler-90 g with ground-based observatories in May 2024. We then combined the 2024 transit and several previously unpublished transit times of planets g and h with the Kepler photometry and RV data to update the masses and linear ephemerides of the planets, finding masses for g and h of \MgMedian\, and \MhMedian\, respectively from a Markov Chain Monte Carlo analysis. These results enable further insights into the architecturally rich Kepler-90 system and pave the way for atmospheric characterization with space-based facilities.

\end{abstract}


\section{Introduction}

Thousands of extra-solar planets have been discovered over the last few decades, many of which belong to multi-planet transiting systems discovered by the NASA Kepler mission (\cite{borucki2010}, Table \ref{table:multiplicity-table}). These high-multiplicity systems illustrate the dynamical interactions and architectural patterns that naturally occur, informing planetary formation models and placing our own solar system into a larger context. To date, only one other star is known to host as many planets as our own sun, namely Kepler-90\footnote{Alternate names for Kepler-90 are: KOI-351, KIC 11442793, TIC 267667295, GAIA DR2/3 2132193431285570304, and 2MASS J18574403+4918185.} (K90). All of K90's known planets are transiting and were discovered in the Kepler photometry.  Seven of K90's planets (b, c, d, e, f, g, h) were validated by \cite{cabrera2014}, \cite{planet-hunters} and  \cite{lissauer2014}. The eighth planet K90i was announced by \cite{planet8} to be transiting with a period of 14.4 days, placing it between planets K90c and K90d as the third-closest planet to its host.

For multi-planet systems like K90, consecutive transits with sufficient photometric precision provide rich dynamical information about the planet orbits, complementing the information about planet radii and mean orbital periods gleaned from individual transits. In particular, the transit timing variation (TTV) due to gravitational interactions between planets can help determine the planet masses and orbital eccentricities \citep{Agol2005,Holman2005}. K90's outer two gas giants, K90g ($P_g$ = 210.7 d, $R_g$ = 8.1 $\pm$ 0.8 $R_\oplus$) and K90h ($P_h$ = 331.6 d, $R_h$ = 11.3 $\pm$ 1.0 $R_\oplus$), have exceptional, strongly-detected TTVs in the Kepler prime mission photometry due to their strong gravitational interactions and large transit depths. The TTVs of K90g and K90h range from several minutes to tens of hours, with the last transit of K90g observed by Kepler in November 2012 occurring 25.7 hours later than expected from the linear ephemeris used in \cite{cabrera2014}.

Planets K90g and K90h, deemed to be dynamically decoupled from the inner planets, were studied in depth by \cite{Liang_2021} using both TTVs and transit duration variations from the Kepler photometry. They found K90g and K90h to have masses of $M_g$ = $15.0^{+0.9}_{-0.8}$ $M_\oplus$ and $M_h$ = $203^{+5}_{-5}$ $M_\oplus$, respectively. With its large radius (Saturn-sized) and relatively low mass (Neptune mass), K90g appears to have an especially low density, consistent with a class of planets dubbed ``super puffs'' \citep{chachan2020}.  K90h's radius and mass are similar to those of Jupiter. \cite{Liang_2021} also affirmed the stability of the K90 system given their determination of K90g and K90h's properties.

In this work, we leverage a decade of radial velocities (RVs) of K90 from the W. M. Keck Observatory to determine the most accurate masses and orbital properties yet for K90 g and h \citep{Weiss2024}.  The RVs have several characteristics that distinguish them from the TTVs: (1) they span a decade, which is substantially longer than the 4-year Kepler prime mission, (2) they provide dynamical information at times between transits, (3) they probe the dynamics of a distinct body (the star, rather than the planets). These attributes make a decade-long RV data set valuable and complementary to the photometry that yielded several precise, but discrete, TTVs.  By analyzing the RVs and TTVs, both separately and jointly, we can gain new insights about the masses and orbits of the planets, as well as the relative information contributed by RVs and TTVs.

Furthermore, as part of this study, we observed an additional transit of K90g in May 2024 from a suite of ground-based facilities.  We include that full transit and several other previously unpublished transits in our analysis to refine the orbital ephemerides and masses of K90g and K90h.
 
Our paper is structured as follows. In \S\ref{sec:measurements} we review the RV data from the Keck Observatory and present newly derived transit times for planets K90g and K90h. In \S\ref{sec:jointly-modeleling-rvs-and-ttvs}, we describe our joint RV-TTV model and fit it first to the Keck RV and Kepler Transit data. We then present our May 2024 transit observation of K90g in \S\ref{sec:2024-observation}, which was added to our dataset along with several other post-Kepler transits of K90g and K90h to allow for a revised fit with our joint RV-TTV model. This updated analysis is described in \S\ref{sec:2024-re-analysis}. Finally, in \S\ref{sec:discussion} we examine the unique contributions of the new RV dataset and the updated TTV dataset, draw comparisons between Kepler-90 and our own solar system, and assess how the ephemeris uncertainty impacts our ability to forecast future transits of K90g and K90h.


\begin{deluxetable}{cc}

\label{table:multiplicity-table}

\tablecaption{Confirmed Planetary System Multiplicities}

\tablehead{\colhead{Multiplicity} & \colhead{Number of Systems}} 

\startdata
1 & 3325 \\
2 & 639 \\
3 & 209 \\
4 & 75 \\
5 & 29 \\
6 & 11 \\
7 & 1 \\
8 & 2 \\
\enddata
\tablecomments{Multiplicities are from the NASA Exoplanet Archive \citep{exo-archive}, accessed July 30, 2024. The two systems with multiplicity 8 are the solar system and Kepler-90.}
\end{deluxetable}

\label{sec:introduction}

\section{Observations} \label{sec:measurements}

\subsection{Transit Timing Variations} \label{subsec:ttv-measurements}

Eight planets have been identified in the Kepler photometry to transit Kepler-90, but in this work we only consider the two outermost planets: K90g ($P_g \approx 211$ d, six transits in Kepler photometry) and K90h ($P_h \approx 332$ d, three transits in Kepler photometry). We used the transit timing variations (TTVs) derived from the Kepler dataset in our initial analysis (see \S\ref{sec:jointly-modeleling-rvs-and-ttvs}; \citealt{rowe2015}, with Q1-Q17 updates provided via private communication). The results of our analysis were used to predict a transit of K90g in May 2024 which was successfully observed by ground-based facilities (see \S\ref{sec:2024-observation}), adding another transit to the dataset.

There have been several other recovered transits of planets g and h following the Kepler Prime mission. We observed single transits of K90h and K90g in December 2013 and August 2014 respectively with the Spitzer Space Telescope’s InfraRed Array Camera\footnote{The computer on which the photometry was originally analyzed was stolen, making it difficult to trace and describe how this analysis was conducted. We recovered the reported time from archival email correspondence and have conservatively inflated the error bars by a factor of two.}. The SWIFT-Gehrels Ultraviolet/Optical Telescope observed a transit of K90g in January 2014 which we also include in our work (Brett Morris, priv. comm.). A partial transit of K90h was observed with the Palomar Observatory (Shreyas Vissapragada, priv. comm.). Transits of K90g were searched for in the Transiting Exoplanet Survey Satellite (TESS) photometry by \cite{tess-non-detection}, but the one expected transit signal (in sector 40) was not detected, and the TESS observing window did not coincide with a transit of K90h. TESS photometry from dates later than Sector 41 was not examined in \cite{tess-non-detection} and is not included in our analysis.  We tabulate all transit measurements used in this work in Table \ref{table:ttv-data-table}. Note that we index transits beginning at zero which may differ from some sources in the literature.


\begin{deluxetable}{lcccc}

\label{table:ttv-data-table}

\tablecaption{Transits of K90g and K90h used in this work}

\tablehead{\colhead{Planet \& Epoch} & \colhead{Transit Time ($BJD_{TDB}$)} & \colhead{Observatory}}

\startdata
g 0 & 2454980.1013 $\pm$ 0.0015 & Kepler \\
g 1 & 2455190.5512 $\pm$ 0.0017 & Kepler \\
g 2 & 2455401.2858 $\pm$ 0.0018 & Kepler \\
g 4 & 2455822.4796 $\pm$ 0.0012 & Kepler \\
g 5 & 2456033.0663 $\pm$ 0.0014 & Kepler \\
g 6 & 2456244.7389 $\pm$ 0.0013 & Kepler \\
g 8 & 2456665.64 $\pm$ 0.11 & Swift$^A$ \\
g 9 & 2456877.007 $\pm$ 0.015 & Spitzer$^B$ \\
g 26 & 2460458.7241 $\pm$ 0.0033 & Ground-Based$^B$ \\
\hline
h 0 & 2454973.4787 $\pm$ 0.0011 & Kepler \\
h 1 & 2455305.12063 $\pm$ 0.00092 & Kepler \\
h 4 & 2456299.89279 $\pm$ 0.00088 & Kepler \\
h 5 & 2456631.505 $\pm$ 0.026 & Spitzer$^B$ \\
\enddata

\tablecomments{Kepler transit times are from Jason Rowe (priv. comm), based on the method of \citet{rowe2015}.  A -- Swift transit time from Brett Moris, priv. comm.  B -- This work.}
\end{deluxetable}


\subsection{Radial Velocities} \label{subsec:rv-measurements}

As a part of the Kepler Giant Planet Search, \cite{Weiss2024} collected 34 radial velocity (RV) measurements of Kepler-90 on the High Resolution Echelle Spectrometer (HIRES, \citealt{Vogt1994}) at the W. M. Keck Observatory between April 2011 and June 2022. We present those RVs in Table \ref{table:rv-data-table}. These are the first published RV measurements of Kepler-90, allowing for both an independent analysis of the system based on RV data as well as a joint analysis of the combined TTV and RV datasets. Because there are so few transit observations of planets g and h, the addition of these RV measurements allows for a substantial improvement in our ability to glean information about these two planets in particular and the K90 system more generally.


\begin{deluxetable}{ccc}

\label{table:rv-data-table}

\tablecaption{Kepler-90 RVs}

\tablehead{\colhead{RV Number} & \colhead{Time} & \colhead{RV} \\ 
\colhead{-} & \colhead{($BJD_{TDB}$)} & \colhead{(m/s)} } 

\startdata
1 & 2455669.105611 & -15.4 $\pm$ 7.3 \\
2 & 2456192.763738 & 13.8 $\pm$ 5.3 \\
3 & 2456194.876885 & 15.0 $\pm$ 7.7 \\
4 & 2456195.835041 & 10.8 $\pm$ 5.6 \\
5 & 2456196.770269 & 31.0 $\pm$ 5.7 \\
6 & 2456207.775114 & 33.7 $\pm$ 6.3 \\
7 & 2456208.734969 & 25.6 $\pm$ 5.8 \\
8 & 2456449.879634 & -27.7 $\pm$ 5.7 \\
9 & 2456468.925956 & -2.3 $\pm$ 6.9 \\
10 & 2456473.032607 & -8.9 $\pm$ 5.8 \\
11 & 2456522.008180 & 8.0 $\pm$ 6.3 \\
12 & 2456532.891596 & 10.0 $\pm$ 5.7 \\
13 & 2456588.767279 & -1.7 $\pm$ 5.5 \\
14 & 2456709.156169 & -2.4 $\pm$ 6.5 \\
15 & 2456845.013523 & 24.6 $\pm$ 5.3 \\
16 & 2456855.909698 & 21.0 $\pm$ 6.1 \\
17 & 2456862.908851 & 7.7 $\pm$ 6.0 \\
18 & 2456881.834946 & 17.1 $\pm$ 5.9 \\
19 & 2456891.850418 & -0.3 $\pm$ 6.2 \\
20 & 2456911.940306 & 11.9 $\pm$ 6.7 \\
21 & 2457002.701540 & -10.5 $\pm$ 7.9 \\
22 & 2457151.091032 & 6.4 $\pm$ 5.7 \\
23 & 2457326.765069 & -24.5 $\pm$ 5.9 \\
24 & 2457583.999116 & -0.7 $\pm$ 5.4 \\
25 & 2457964.010530 & -10.6 $\pm$ 4.6 \\
26 & 2457988.826036 & -25.6 $\pm$ 5.0 \\
27 & 2458000.929813 & -18.3 $\pm$ 5.8 \\
28 & 2458018.840170 & -14.9 $\pm$ 4.7 \\
29 & 2458091.706253 & -28.7 $\pm$ 5.2 \\
30 & 2459034.888522 & -29.3 $\pm$ 4.8 \\
31 & 2459377.886027 & -13.9 $\pm$ 4.7 \\
32 & 2459413.018778 & -6.6 $\pm$ 4.9 \\
33 & 2459484.847952 & 21.4 $\pm$ 4.3 \\
34 & 2459745.078442 & -19.8 $\pm$ 5.2 \\
\enddata

\tablecomments{RVs are from \cite{Weiss2024}.}

\end{deluxetable}

\section{Jointly Modeling RVs and Kepler TTVs} \label{sec:jointly-modeleling-rvs-and-ttvs}

Although Kepler-90 is an eight-planet system, we only model the outer two planets, K90g and K90h, in keeping with a previous approach \citep{Liang_2021}.  This choice keeps the model computationally tractable, with only 12 free parameters (five for each planet plus two nuisance parameters), as opposed to 42 free parameters for a full Keplerian model of eight planets. Our choice to simplify the model is based on the comparatively low-signal, noisy TTV and RV data for the small, inner planets, whereas the outer two planets are large, have significant TTV signals, and are expected to yield significant RV signals. In this section, we only consider the RVs and TTVs from Kepler, leaving an analysis of the complete dataset for \S\ref{sec:2024-re-analysis}.

\subsection{Optimization Procedure} \label{subsec:optimization}

\begin{figure*}[ht!]
\plotone{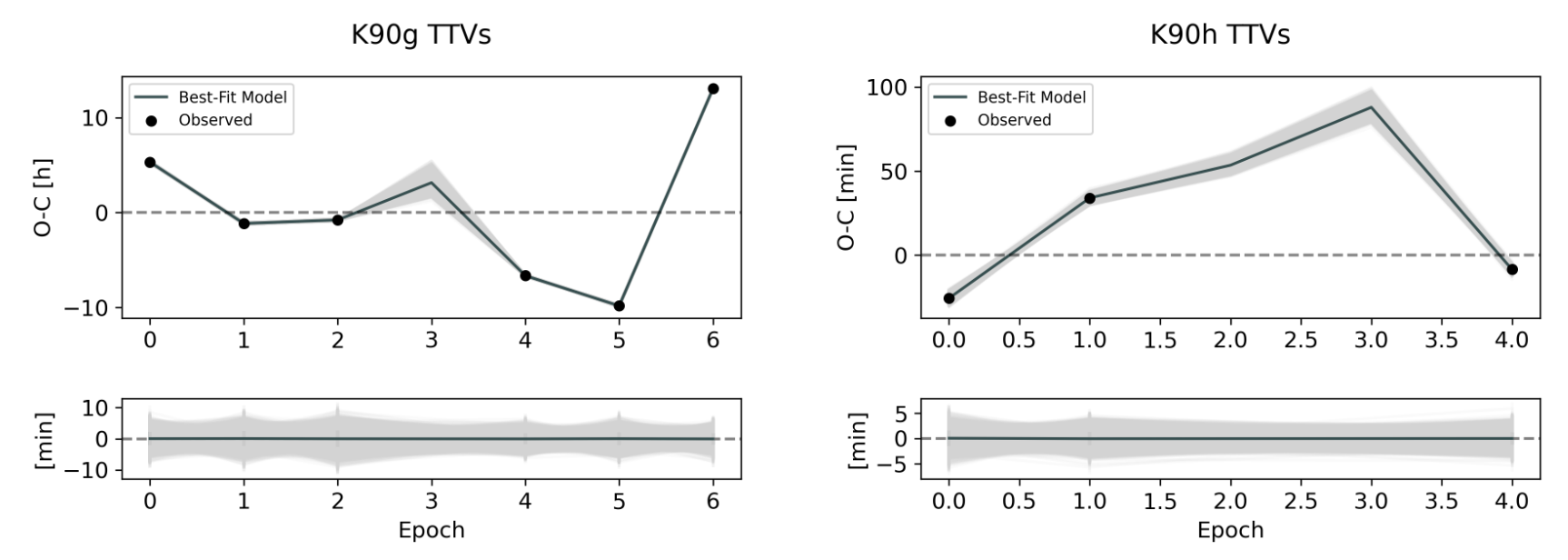}
\caption{(Tops) Observed TTVs (O-C) of planets g (left) and h (right) from the Kepler photometry along with the best-fit model from our joint analysis of the RV and Kepler TTV data (black line) and the 6438 models from the Monte Carlo estimate of the posterior distribution (gray lines). The linear ephemeris used to determine these O-C values was determined by performing a linear regression on the observed Kepler transit times. (Bottoms) Residuals after subtracting the modeled transit times from the observed transit times.}
\label{fig:ttv-data-joint-fits-g-and-h-plots}
\end{figure*}

\begin{figure}[ht!]
\plotone{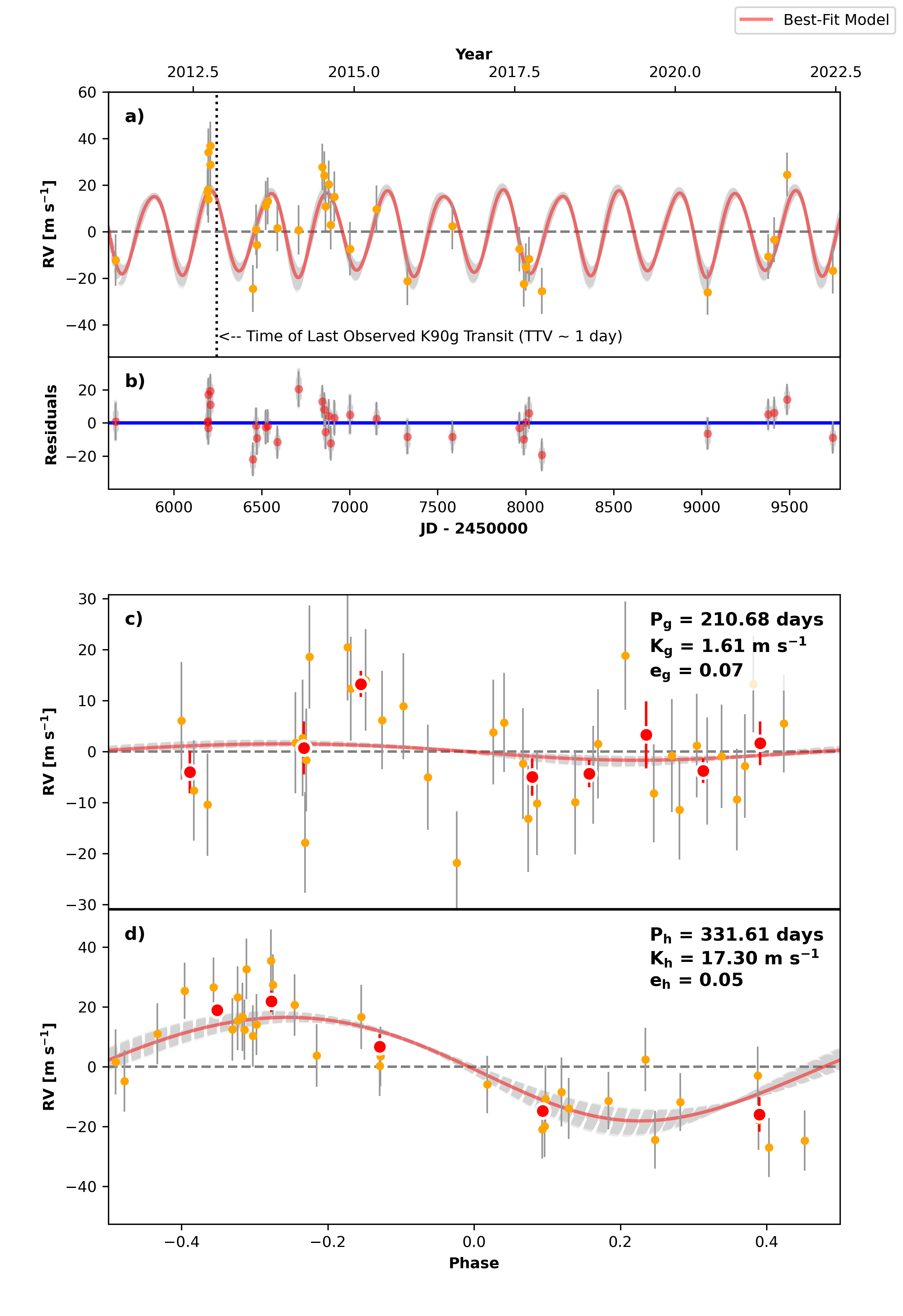}
\caption{(a) Observed RVs for K90 from the HIRES spectrometer along with the best-fit model from our joint analysis of the RV and Kepler TTV data (red line) and the 6438 models from the Monte Carlo estimate of the posterior distribution (gray dashed lines). The time of the last observed transit from the Kepler photometry is indicated (dotted black line).  Panel (b) shows the RV residuals from the joint model. Panels (c) and (d) show the RV data phased to the periods of K90g and K90h along with the best-fit model and posteriors as in panel (a). The period, semi-amplitude (K), and eccentricities correspond to the best fit to the joint RV and Kepler TTV model.}
\label{fig:rv-data-and-model-spread-joint-analysis-plot}
\end{figure}

We simultaneously modeled the individual transit times of K90g and K90h from the Kepler dataset (Fig. \ref{fig:ttv-data-joint-fits-g-and-h-plots}) along with the RVs of Kepler-90 (Fig. \ref{fig:rv-data-and-model-spread-joint-analysis-plot}). We forward modeled the transit times of planets g and h using the Python implementation of TTVFaster\footnote{We also modeled the transit times and RVs with the N-body integrator TTVFast and found masses for K90g and K90h consistent with our analysis presented here \citep{TTVFast}.} \citep{ttvFaster}. This code employs an efficient analytical method of computing transit times for planets (given their average linear ephemeris) to first order in planet-star mass ratios and in the orbital eccentricities. We chose this model for its simplicity and speed compared to traditional N-body integrators and photodynamical models. In using TTVFaster, the planets’ orbits are assumed to be co-planar and edge-on, approximations which are suitable for these transiting planets and are consistent with the values determined by \citealt{Liang_2021} (see also \citealt{fabrycky2014}, \citealt{yee2021}).
We used RadVel: The Radial Velocity Fitting Toolkit \citep{radvel} to solve Kepler's equation and forward model the expected Kepler-90 RVs.

Our joint model has as free parameters the mass (m), period (P), eccentricity and argument of periastron (as $\sqrt{e}cos(\omega)$ and $\sqrt{e}sin(\omega)$), and linearly fit time of conjunction (T0) for each planet, as well as two nuisance parameters: the HIRES RV offset ($\gamma$) and jitter ($\sigjit$).  We initially fixed the stellar mass of Kepler-90 to the central value found by \cite{cabrera2014} of 1.2 solar masses as in \cite{Liang_2021} (although we later allowed the stellar mass to vary in \S\ref{sec:2024-re-analysis}). We used uninformative priors on the eccentricities ($0 < e < 1$), RV zero-point ($-50 < \gamma < 50$) and RV jitter ($0 < \sigjit < 30$). We enforced the masses of g and h to be positive, and we rejected parameter sets which led to orbit crossing of planets g and h.

We optimized our model using the following log-likelihood containing contributions from both Transit Times (TTs) and RVs:

\begin{eqnarray}
    \text{ln}\mathcal{L} &=& -\frac{1}{2}\left(\chi_{RV}^{2} + \Sigma_{i=1}^{N_{RV}}\text{ln}\left[2\pi(\sigma_{RV,i}^{2}+\sigjit^{2})\right]\right) \cr 
    &-& \frac{1}{2}\left(\chi_{TT}^{2} + \Sigma_{i=1}^{N_{TT}}\text{ln}\left[2\pi\sigma_{TT,i}^{2}\right]\right),
\label{eqn:lnlike}
\end{eqnarray}
where $\sigma_{RV,i}$ and $\sigma_{TT,i}$ are the measurement uncertainties in the $i$th RV and $i$th transit time measurements, respectively, $\sigjit$ is the RV error induced by stellar activity and systematics, and $\chi^2$ refers to the Chi-Squared statistic, defined for each type of data:
\begin{equation}
    \chi_{RV}^{2} = \sum_{i=1}^{N_{RV}}\frac{\left(RV_{\mathrm{obs},i}-RV_{\mathrm{mod},i}\right)^{2}}{\sigma_{RV,i}^{2}+\sigjit^{2}},
\label{eqn:chi2-rv}
\end{equation}
\begin{equation}
    \chi_{TT}^{2} = \sum_{i=1}^{N_{TT}}\frac{\left(TT_{\mathrm{obs},i}-TT_{\mathrm{mod},i}\right)^{2}}{\sigma_{TT,i}^{2}}.
\label{eqn:chi2-tt}
\end{equation}
The sum over transit observations includes both K90g and K90h transits. Note that the last term of Equation \ref{eqn:lnlike}, $-\Sigma_{i=1}^{N_{TT}}\frac{1}{2}\text{ln}(2\pi\sigma_{TT,i}^{2})$, is a constant offset fully determined by the observational error and does not depend on our free parameters. It is included for completeness. We emphasize that these equations involve the measured times of transit, not the O-C values derived from subtracting a linear ephemeris.

We added the log-likelihood and log-prior to form the (unnormalized) posterior log-probability, and we minimized the negative of this quantity using sequential least squares programming (SLSQP) as implemented in the Python package SciPy \citep{2020SciPy-NMeth}. The best-fit parameters are listed in Table \ref{table:joint-results-table}, and the model's fit to the data is shown in Figs. \ref{fig:ttv-data-joint-fits-g-and-h-plots} (for the transit times) and \ref{fig:rv-data-and-model-spread-joint-analysis-plot} (for the RVs).

\subsection{Uncertainty Estimation} \label{subsec:uncertainties}

\begin{figure}[ht!]
\plotone{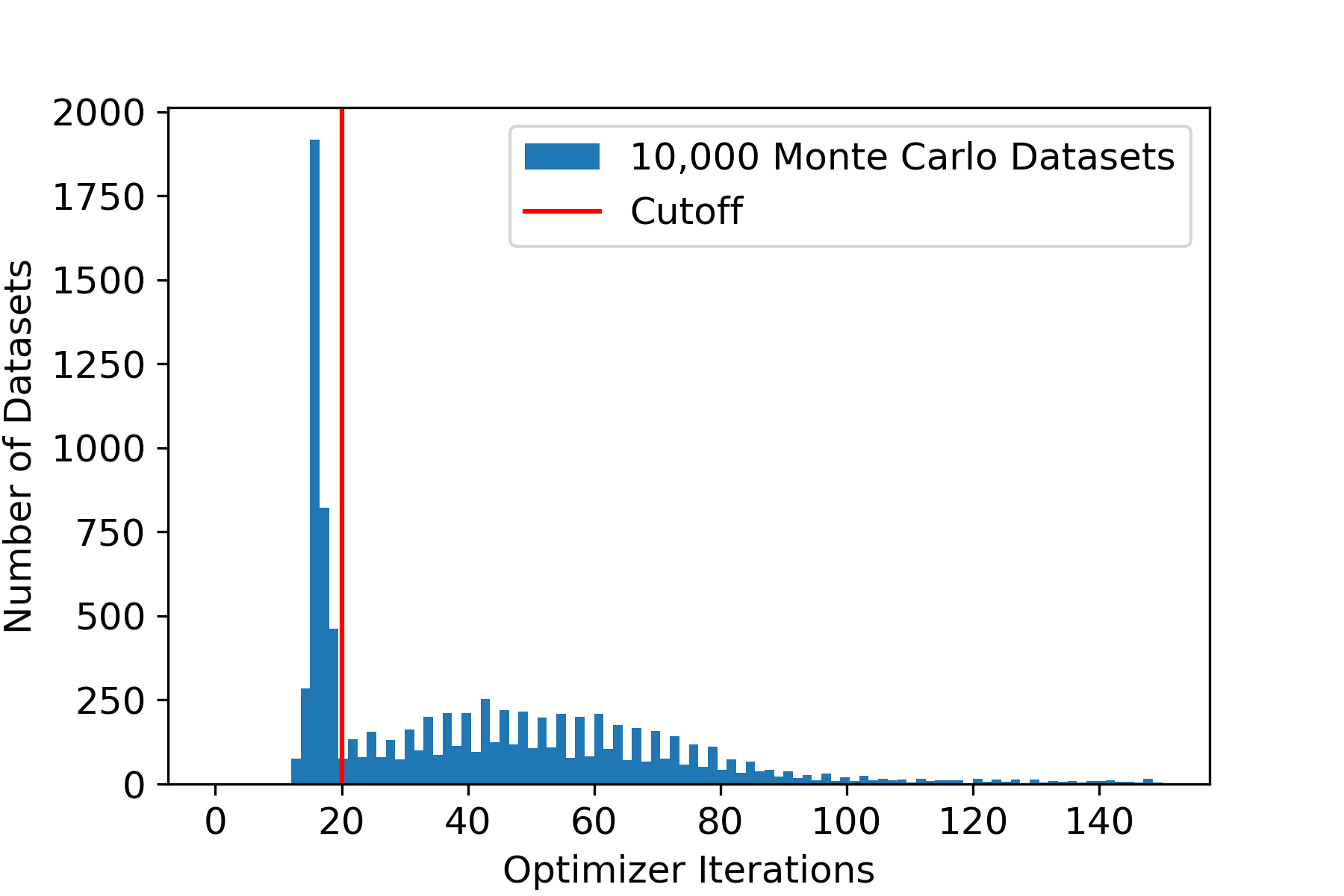}
\caption{Histogram of the number of iterations performed by the optimizer from 10,000 Monte Carlo synthetic TTV and RV datasets (blue). We removed synthetic datasets for which the optimizer stopped in less than 20 iterations (red line), resulting in a final set of 6438 optimizations.}
\label{fig:iteration-hist-plot}
\end{figure}

We first attempted to explore the posterior distribution using a Markov chain Monte Carlo (MCMC) analysis via the Python package emcee3 \citep{emcee}. However, the chains did not converge after 750,000 steps with 128 walkers as determined by the integrated autocorrelation time \footnote{Note that with the inclusion of post-Kepler transits, this non-convergence was no longer a problem. \S\ref{sec:2024-re-analysis} presents a convergent MCMC analysis of the complete dataset.}. To circumvent this non-convergence, we explored confidence intervals using what is sometimes called a ``bootstrap'' Monte Carlo simulation, as follows (see e.g. \cite{numerical-recipes} \S\ 15.6). Let $\mathcal{D}_0$ denote the set of measurements (RVs and Kepler's transit midpoints) and $\bold{a_0}$ denote the best-fit solution to our joint model found by the optimization procedure in \S\ref{subsec:optimization}. We parameterized each measurement from the original $\mathcal{D}_0$ by a normal distribution with mean value equal to the measured value and standard deviation equal to the measurement uncertainty,  $\mathcal{N}(x_{obs},\sigma_{obs})$. We generated 10,000 synthetic datasets $\mathcal{D}_i$ by drawing from $\mathcal{N}(x_{obs},\sigma_{obs})$. For each  $\mathcal{D}_i$, we optimized our model according to 
\S\ref{subsec:optimization}, resulting in a set of best-fit parameters $\{\bold{a_i}\}$. The initial guess used for each synthetic optimization was the best-fit to $\mathcal{D}_0$. One disadvantage of this approach is that the optimizer sometimes settled into a minimum near the starting value after only a few iterations, which could lead to underestimated uncertainties. To counteract this, we discarded any optimization run that terminated in fewer than 20 iterations. We chose $<20$ iterations as the threshold for removal based on a large pileup in the optimizer iteration histogram (Fig. \ref{fig:iteration-hist-plot}). After discarding the dubious solutions, 6438 $\bold{a_i}$ parameter vectors remained. We then made the assumption that the distribution $\bold{a_i}-\bold{a_0}$ is approximately the same as the distribution $\bold{a_i}-\bold{a_{true}}$, where $\bold{a_{true}}$ represents the true values of our model parameters. We could therefore approximate the underlying distributions of and derive uncertainties for our parameter estimates (Table \ref{table:joint-results-table}). This Monte Carlo method results in median masses of $M_g = 17.6^{+1.8}_{-1.5}$ $M_\oplus$ and $M_h = 215^{+9.8}_{-7.4}$ $M_\oplus$.

\begin{figure*}[ht!]
\plotone{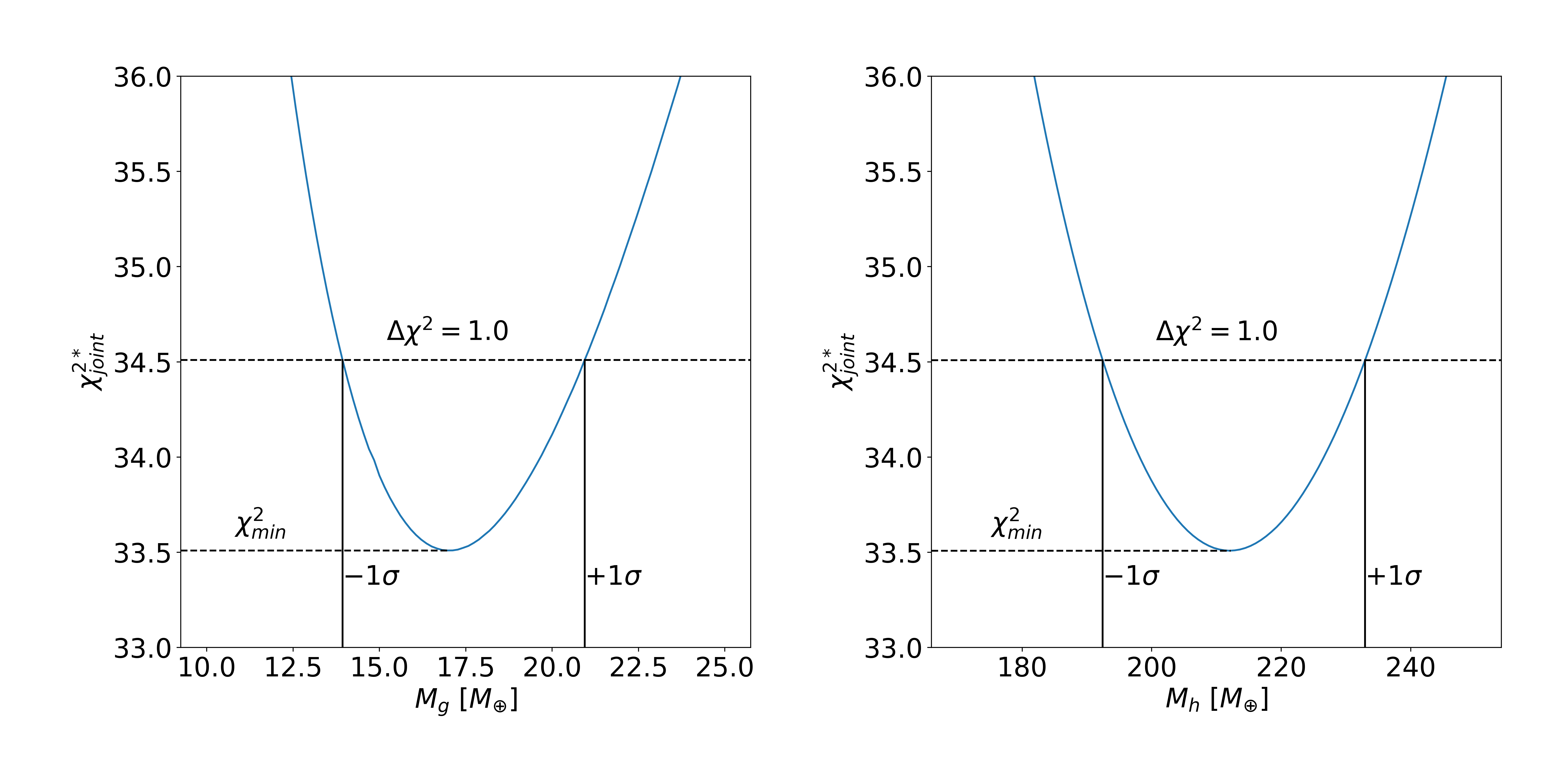}
\caption{Minimum RV and Kepler TTV $\chi^{2*}_{\mathrm{joint}}$ values (Eqs. \ref{eqn:chi2-rv} $+$ \ref{eqn:chi2-tt} with fixed best-fit $\sigjit$) projected onto the 1D parameter subspaces for the masses of K90g (left) and K90h (right). In each case, the mass was held fixed at a given value while $\chi^{2}_{\mathrm{joint}}$ was minimized as described in \S\ref{subsec:uncertainties}. The resulting change in $\chi^{2*}_{joint}$ above the minimum approximately follows a $\chi^2$ distribution with one degree of freedom, giving 1$\sigma$ confidence limits on the masses of planets g ($M_g = 17.0^{+3.9}_{-3.1}$ $M_\oplus$) and h ($M_h = 212^{+21}_{-20}$ $M_\oplus$).}
\label{fig:chi2-confidence-plots}
\end{figure*}

We performed an approximate consistency check on the mass estimates for K90g and K90h by examining constant $\chi^2$ boundaries for the two parameters as follows (see e.g. \citealt{numerical-recipes}). We first fixed $\sigjit$ to the best-fit value so that our model is effectively reduced to have 11 free parameters. We then varied one mass value at a time over a range around its corresponding best-fit value: (10-25 $M_\oplus$ for g and 170-250 $M_\oplus$ for h). For each mass of a given planet we minimized $\chi^2_{\mathrm{joint}} = \chi^2_{TT} + \chi^2_{RV}$ using the same SLSQP optimizer described previously. Denote this minimum value by $\chi^{2*}_{\mathrm{joint}}$. This collection of $\chi^{2*}_{\mathrm{joint}}$ values over the range of masses for $M_g$ ($M_h$) traces out a path in the 11-dimensional parameter space which projects down onto the one-dimensional subspace of whichever mass was scanned over.
By assuming normally distributed measurement errors and treating our model as linearized in the small region considered near the best-fit value, the variable $\Delta \chi^2 \equiv \chi^{2*}_{\mathrm{joint}} - \chi^2_{\mathrm{min}}$ is approximately distributed as a chi-square distribution with one degree of freedom. Thus, our $1\sigma$ confidence limits on the masses of planets g and h are those values at which $\Delta \chi^2 = 1$. This $\chi^2$ profile analysis results in $M_g = 17.0^{+3.9}_{-3.1}$ $M_\oplus$ and $M_h = 212^{+21}_{-20}$ $M_\oplus$, and the graphs of $\chi^{2*}_{\mathrm{joint}}$ can be seen in Fig. \ref{fig:chi2-confidence-plots}. These errors are approximately twice as large as the errors deduced from our bootstrap Monte Carlo analysis (Table \ref{table:joint-results-table}). This discrepancy is likely because one or more of our above assumptions does not hold, but with the addition of post-Kepler transit data (especially the 2024 observation presented in \S\ref{sec:2024-observation}) we arrive at a more robust error analysis in \S\ref{sec:2024-re-analysis}.

\begin{deluxetable*}{ccccc}

\label{table:joint-results-table}

\tablecaption{Joint RV-TTV Analysis Results}

\tablehead{\colhead{Model Parameter} & \colhead{Best Fit} & \colhead{Monte Carlo} & \colhead{Best Fit} & \colhead{\textbf{MCMC}} \\
\colhead{} & \colhead{(RV + Kepler)} & \colhead{(RV + Kepler)} & \colhead{(RV + All Transits)} & \colhead{\textbf{(RV + All Transits)}} }

\startdata
$M_g$ [$M_\oplus$]        & 16.9     & 17.6$^{+1.8}_{-1.5}$           & 14.8      & 15.0$\pm 1.3$ \\ 
$M_h$ [$M_\oplus$]        & 211.2    & 215.0$^{+9.8}_{-7.4}$          & 201       & 203$\pm 16$ \\ 
$P_g$ [days]              & 210.681  & 210.667$^{+0.026}_{-0.033}$    & 210.73514 & 210.73514$\pm 0.00017$ \\ 
$P_h$ [days]              & 331.6107 & 331.6128$^{+0.0055}_{-0.0040}$ & 331.60296 & 331.60296$\pm 0.00034$ \\ 
$T0_g$ [BJD-2454900]      & 80.14    & 80.18$^{+0.10}_{-0.08}$        & 79.97603  & 79.97608$^{+0.00091}_{-0.00092}$\\
$T0_h$ [BJD-2454900]      & 73.485   & 73.481$^{+0.008}_{-0.011}$     & 73.50058  & 73.50056$^{+0.00084}_{-0.00081}$\\
$\sqrt{e_g}cos{\omega_g}$ & -0.233   & -0.257$^{+0.050}_{-0.049}$     & -0.073    & -0.071$^{+0.012}_{-0.011}$\\
$\sqrt{e_g}sin{\omega_g}$ & 0.109    & 0.098$^{+0.025}_{-0.022}$      & 0.158     & 0.155$^{+0.010}_{-0.011}$\\
$\sqrt{e_h}cos{\omega_h}$ & -0.200   & -0.219$\pm 0.040$              & -0.0820   & -0.0805$^{+0.0098}_{-0.0091}$\\
$\sqrt{e_h}sin{\omega_h}$ & 0.120    & 0.112$^{+0.017}_{-0.016}$      & 0.1471    & 0.1450$^{+0.0085}_{-0.0093}$\\
$\gamma$ [m/s]            & -3.14    & -3.17$^{+0.98}_{-0.94}$        & -3.0      & -3.0$\pm 1.8$ \\ 
$\sigjit$ [m/s]   & 8.4      & 9.9$\pm 1.0$                   & 8.5       & 8.9$^{+1.7}_{-1.5}$\\
$M_{star}$ [$M_\odot$]    & Fixed at 1.2        & Fixed at 1.2        & 1.227     & 1.242$^{+0.096}_{-0.097}$\\
\hline
Derived Parameter         &          &                                &   & \\
\hline
$e_g$                     & 0.066    & 0.076$^{+0.024}_{-0.019}$      & 0.0302 & 0.0292$\pm 0.0038$ \\ 
$e_h$                     & 0.054    & 0.061$^{+0.016}_{-0.013}$      & 0.0284 & 0.0276$\pm 0.0031$ \\ 
$\omega_g [deg]$          & 154.9    & 159.0$^{+6.8}_{-9.4}$          & 114.9 & 114.6$^{+3.5}_{-3.8}$ \\
$\omega_h [deg]$          & 149.1    & 152.9$^{+6.7}_{-8.4}$          & 119.1 & 119.0$^{+3.0}_{-3.1}$ \\
$\chi^2_{\nu}$                & 1.08 (DoF = 31)   & -                     & 1.27 (DoF = 34) & - \\
\enddata

\tablecomments{$\chi^2_{\nu}$ is the reduced $\chi^2$ statistic, with degrees of freedom (DoF) indicated.}

\end{deluxetable*}


\section{Transit Recovery in 2024}
\label{sec:2024-observation}
\begin{figure*}
\plotone{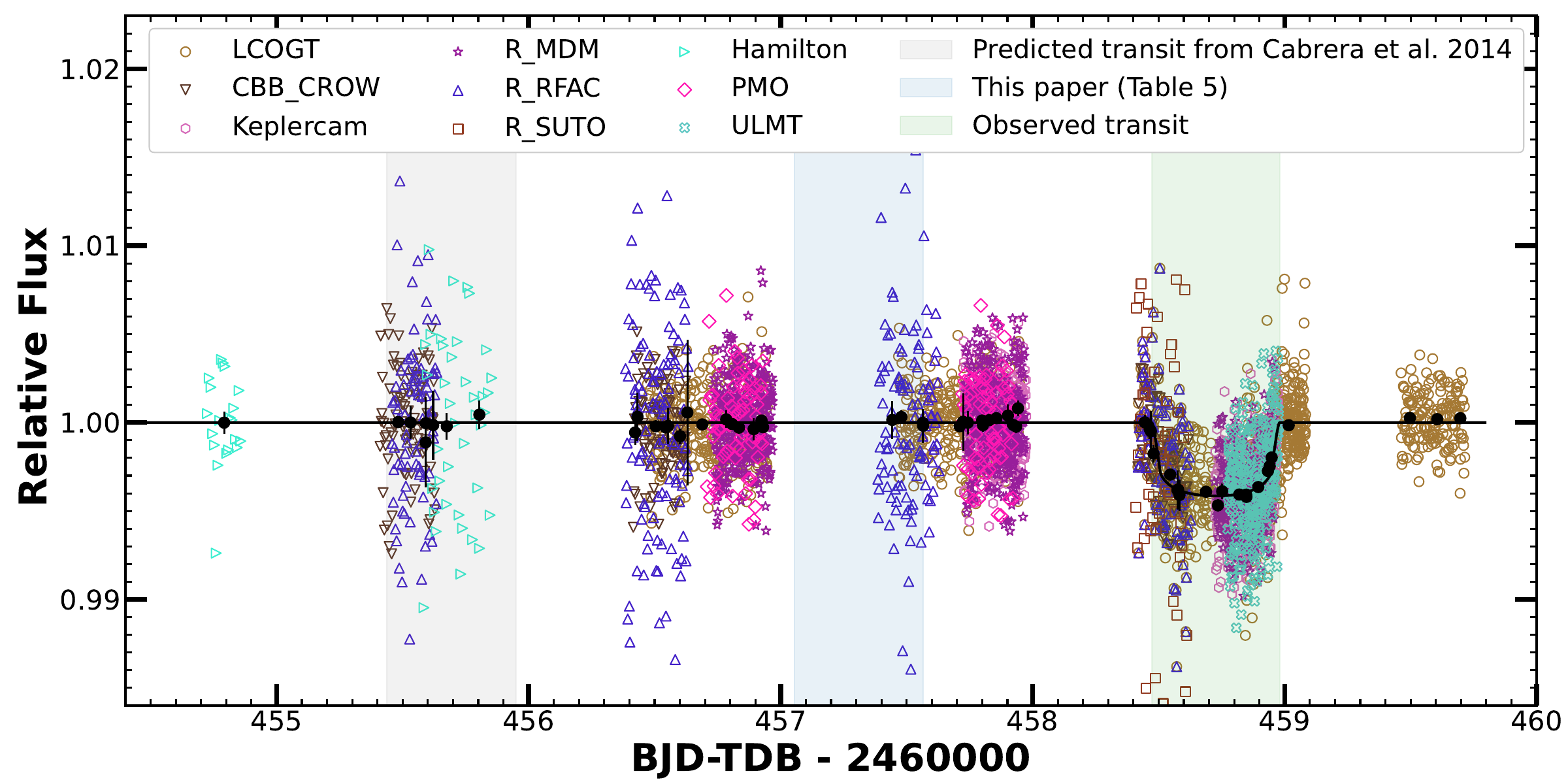}
\caption{Kepler-90 photometry from the SG1 collaboration capturing ingress and egress of K90g between UT 2024-05-25 and 2024-05-29. The preliminary transit midpoint of $2460458.724054^{+0.00164}_{-0.00330}$ BJD-TDB at epoch 26 occurred approximately 1.4 days later than predicted by our model (Fig. \ref{fig:ttv-predictions-plots-kepler}). This transit delay is consistent with other post-Kepler observations in suggesting that our best-fit period for planet g from the RV and Kepler data alone is shorter than the true value. The observations from these various facilities are described in \S\ref{sec:2024-observation}.}
\label{fig:k90g-2024-transit-lightcurve-preliminary}
\end{figure*}

During the preparation of this manuscript, a partial transit of K90g that covered both ingress and egress was observed by multiple ground-based facilities in the TESS Follow-up Observing Program \citep{collins:2018} Sub Group 1 (SG1) using the transit predictions made by our model as presented above (Fig. \ref{fig:k90g-2024-transit-lightcurve-preliminary}). Based on the 95\% confidence spread in transit times, we recommended an observing window spanning five days, centered on our predicted transit midpoint. Because the transit duration is 12 hours, we used an intensive coordinated effort across facilities at various longitudes.  With this strategy, we detected the transit ingress, in-transit, and egress, with multiple facilities contributing to the detection, on UT 2024 May 28.  Photometry from the Las Cumbres Observatory Global Telescope \citep[LCOGT;][]{Brown:2013} 1\,m network nodes at Teide Observatory on the island of Tenerife and McDonald Observatory near Fort Davis, Texas, United States captured pre-ingress and post-egress baseline on UTC 2024 May 26, 27, and 29, and an in-transit shift of the expected $\sim 4$\,ppt transit depth on UTC 2024 May 28. Photometry from the MuSCAT3 multi-band imager \citep{Narita:2020} on the 2\,m Faulkes Telescope North at Haleakala Observatory on Maui, Hawai'i captured the transit egress on UTC 2024 May 28. The 0.61\,m University of Louisville Manner Telescope (ULMT) located at the Steward observatory near Tucson, AZ, KeplerCam on the 1.2\,m telescope at the Fred Lawrence Whipple Observatory located on Mt. Hopkins in southern Arizona, and the 1.3\,m McGraw-Hill telescope at the MDM Observatory on Kitt Peak also caught part of egress on UTC 2024 May 28. Three citizen astronomers independently captured photometry consistent with a 4 ppt ingress at the 0.36\,m telescope at CROW observatory near Portalegre, Portugal, the 0.3m Frustaglia Private Observatory (RFAC) at e-EyE near Fregenal de la Sierra, Spain, and the Silesian University of Technology 1 (SUTO1) 0.3\,m telescope located near Otivar, Spain. 

Additional observations that ruled out an ingress or egress event were conducted by the Hamilton College Observatory 0.51\,m telescope near Clinton, New York on UTC 2024 May 24th and 25th, the Pine Mountain Observatory 0.35\,m telescope near Bend, Oregon on UTC 2024 May 26th and 27th, the 1.3\,m McGraw Hill telescope on UTC 2024 May 26th and 27th, and KeplerCam on UTC 2024 May 27th.

We analyzed the photometry using {\tt AstroImageJ} \citep{Collins:2017}. We analyzed the LCOGT 1\,m multi-night image sets (all observed in Sloan $i'$ band) along with images from the MuSCAT3 $i'$ band channel as a single image set. We extracted 30 comparison stars that are closest in proximity and nearest in brightness to K90 that are common to all images. We then minimized RMS over the 30 comparison star set without detrending, which resulted in 10 comparison stars being selected. The $\sim 4$\,ppt in-transit shift and egress were then obvious. We individually detrended the other light curves using the single best of 7 available detrend parameters, if it was justified by the Bayesian Information Criterion (BIC; \citet{schwarz1978}). We then combined the multi-night LCOGT light curve with the other detrended light curves near the transit, allowing for baseline offsets (due to differing filters and comparison ensembles). 

Finally, we performed a Markov-chain Monte Carlo (MCMC) analysis using  the Metropolis-Hastings \citep{Metropolis_1953,Hastings_1970} algorithm implemented in {\tt TRAFIT}, a revised version of the code described in \cite{Gillon2010AA,Gillon2012,Gillon2014AA}.
The transit light curves  were modeled using the quadratic limb-darkening model of \cite{Mandel2002}. We applied a Gaussian prior distribution on the stellar parameters ($R_\star$, $M_\star$, $\log g_\star$, [Fe/H], and $T_{\rm eff}$). The varied parameters  sampled by our MCMC were: the transit midpoint time ($T_{\rm{conj,g}}$), the orbital period, the total transit duration ($T_{\rm{dur}}$), the transit depth ($\delta$), and the combination of quadratic limb-darkening coefficients $q_1 = (u_1 + u_2)^2$ and $q_2 = 0.5u_1(u_1 + u_2)^{-1}$ \citep{Kipping_2013MNRAS.435.2152K}, where $u_1$ and  $u_2$ are calculated from \cite{Claret_2012AA,Claret_2018AandA}. 
Assuming a circular orbit ($e = 0$), we performed an MCMC analysis composed of three Markov chains with 100,000 steps to infer the transit parameters.  The final values are given in Table \ref{tab:new_transit_params}.

\begin{figure*}[ht!]
\plotone{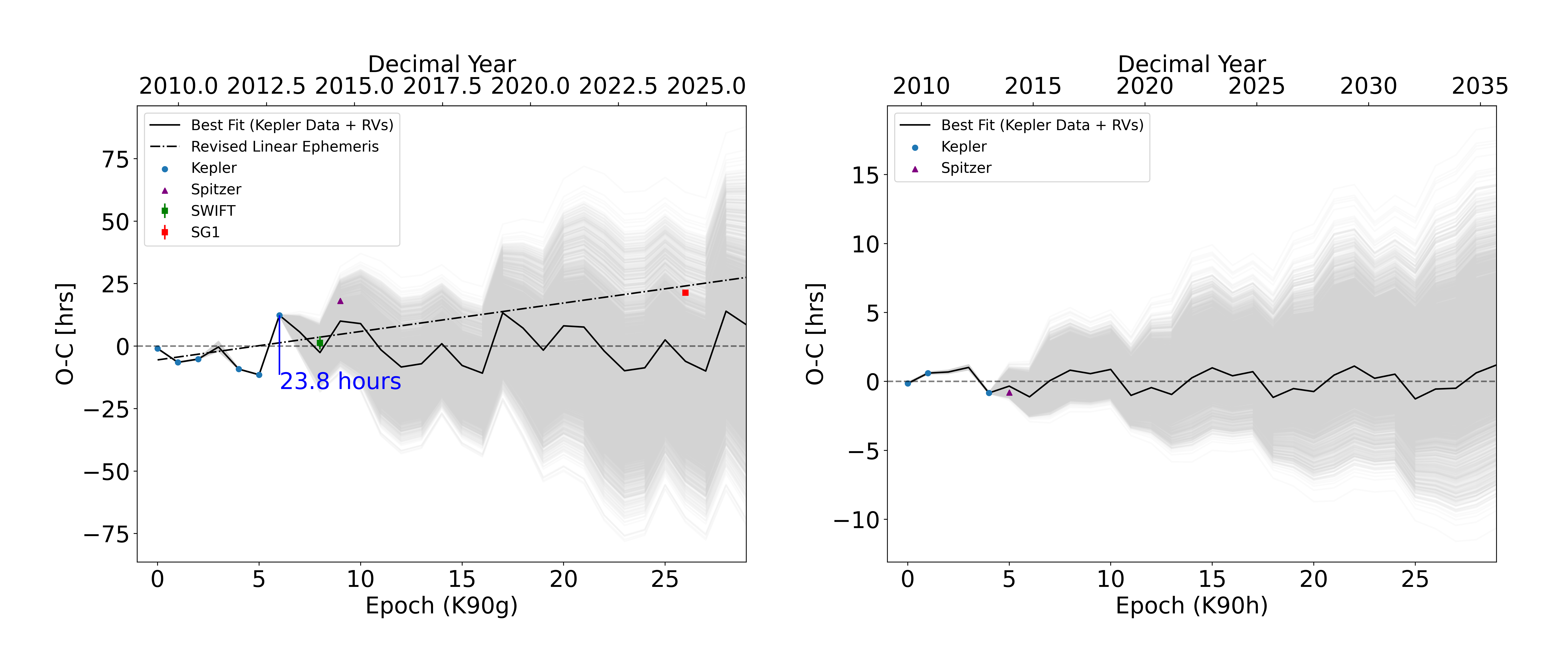}
\caption{Predicted TTVs for planets g (left) and h (right) through 30 transits, starting from the first epoch of the Kepler data and modeled using TTVfaster with the best joint fit parameters (bold line) and the Monte Carlo distribution (gray lines, N=6438). The linear ephemerides used here for each planet are those of the best fit to the joint RV and Kepler TTV model. The four non-Kepler transits reported in this work (see \S\ref{subsec:ttv-measurements}) lie within the range of expected O-C values from our results. The max spread in modeled transit times for the final predicted epoch (29) is 6.65 d for K90g and 1.21 d for K90h. The revised linear ephemeris for planet g including post-Kepler transits (dot-dashed line) shows the preference for a longer period than that suggested by the RV and Kepler data alone.}
\label{fig:ttv-predictions-plots-kepler}
\end{figure*}

Our stand-alone analysis of the 2024 May 28 transit yields a conjunction time of $2460458.724054^{+0.00164}_{-0.00330}$ BJD-TDB \footnote{This conjunction time was determined with a subset of the SG1 data which included the best constraints on ingress and egress. Additional observations not used in the fit were either out-of-transit or partial ingress/egress.}. This recovered transit is indicated in Figure\,\ref{fig:ttv-predictions-plots-kepler} as the red square. The 2024 transit was later than predicted by the linear ephemeris suggested by the RV and Kepler data by approximately 1.4 days (The transits observed between 2012 and 2015 were also later than predicted by this ephemeris).  One interpretation of the increasingly late transits is that the linear ephemeris determined from the Kepler photometry is incorrect, likely because the planet was undergoing significant TTVs during those epochs, and that a slightly longer period is a better fit to the data and a better long-term predictor of planet transits. The unweighted least-squares best fit to all transit times of K90g (from Kepler, Swift, Spitzer, and SG1) suggests a period of 210.7279 days with T0 = 2,454,979.911 BJD compared to the joint Kepler/Hires best-fit of 210.6805 days with T0 = 2,454,980.141 BJD (Figure \ref{fig:ttv-predictions-plots-kepler}). A more refined estimate of the updated linear ephemeris from a combination of all transit and RV data is discussed in \S\ref{sec:2024-re-analysis} below.

\begin{deluxetable}{lcc}
    \centering
    \caption{2024 May Transit Parameters.}
    \label{tab:new_transit_params}
    \tablehead{\colhead{Parameter}   & \colhead{Units}  &  \colhead{Value} ($\pm1\,\sigma$)}
    \startdata
    $T_{\rm{conj,g}}$ & BJD& $2460458.724054^{+0.00164}_{-0.00330}$ \\
    $T_{\rm{dur}}$ & hours & $12.32 \pm 0.18$ \\
    $\delta$ & ppt &  $3.8 \pm 0.1$\\
    $q_1$ & -& $0.313 \pm 0.03$ \\
    $q_2$ & - & $0.244 \pm 0.01$ \\
    $P_g$ & days &  $210.728 \pm 0.017$\\
    \enddata
    \tablecomments{All parameters come from the MCMC fit to the 2024 May photometry, except for $P_g$, which is based on the best-fit linear ephemeris to the transit midpoints of the 2024 May transit, Kepler Q1-Q17, Swift, and Spitzer.}
\end{deluxetable}

\section{Jointly Modeling RVs and All TTVs} \label{sec:2024-re-analysis}

We updated the Kepler-90 planetary parameter estimates by repeating our optimization procedure of \S\ref{sec:jointly-modeleling-rvs-and-ttvs} using the full dataset of RVs and TTVs from Kepler, Swift, Spitzer, and the TFOP SG1 2024 ground-based observations. We used the same models for RVs and transit times, but we allowed the stellar mass of Kepler-90 to vary with a Gaussian prior of $1.2 \pm 0.1 M_\odot$. We again attempted to sample the posteriors with emcee, and with the additional post-Kepler transits there were sufficient data for our MCMC chains to converge (the chains were longer than 50 times the integrated autocorrelation time for each parameter). We initialized 128 walkers in the following way, allowing them to take 10,000 steps each. First, we again used SciPy's SLSQP algorithm to minimize the negative log-probability of our joint RV-TTV model. Next, we initialized the walkers with a standard deviation of 1\% of the best-fit value by adding Gaussian noise from the standard normal distribution. We determined the maximum parameter autocorrelation time, $\tau_{max} \, = \, 191$, then discarded the first $2\tau_{max}$ steps as burn-in and thinned the chains by keeping every $\tau_{max}/2$ steps (rounded down). This procedure left 12,928 posterior samples. The $1 \sigma$ confidence intervals for each parameter are listed in Table \ref{table:joint-results-table}. The parameter posteriors showed no visually noticeable correlations beyond those expected from model degeneracies: $M_g$:$M_h$, $M_g$:$M_{star}$, $M_h$:$M_{star}$, $\sqrt{e_g}cos(\omega_g)$:$\sqrt{e_h}cos(\omega_h)$, $\sqrt{e_g}sin(\omega_g)$:$\sqrt{e_h}sin(\omega_h)$, and P:T0 for g and h individually.

\begin{figure*}[ht!]
\plotone{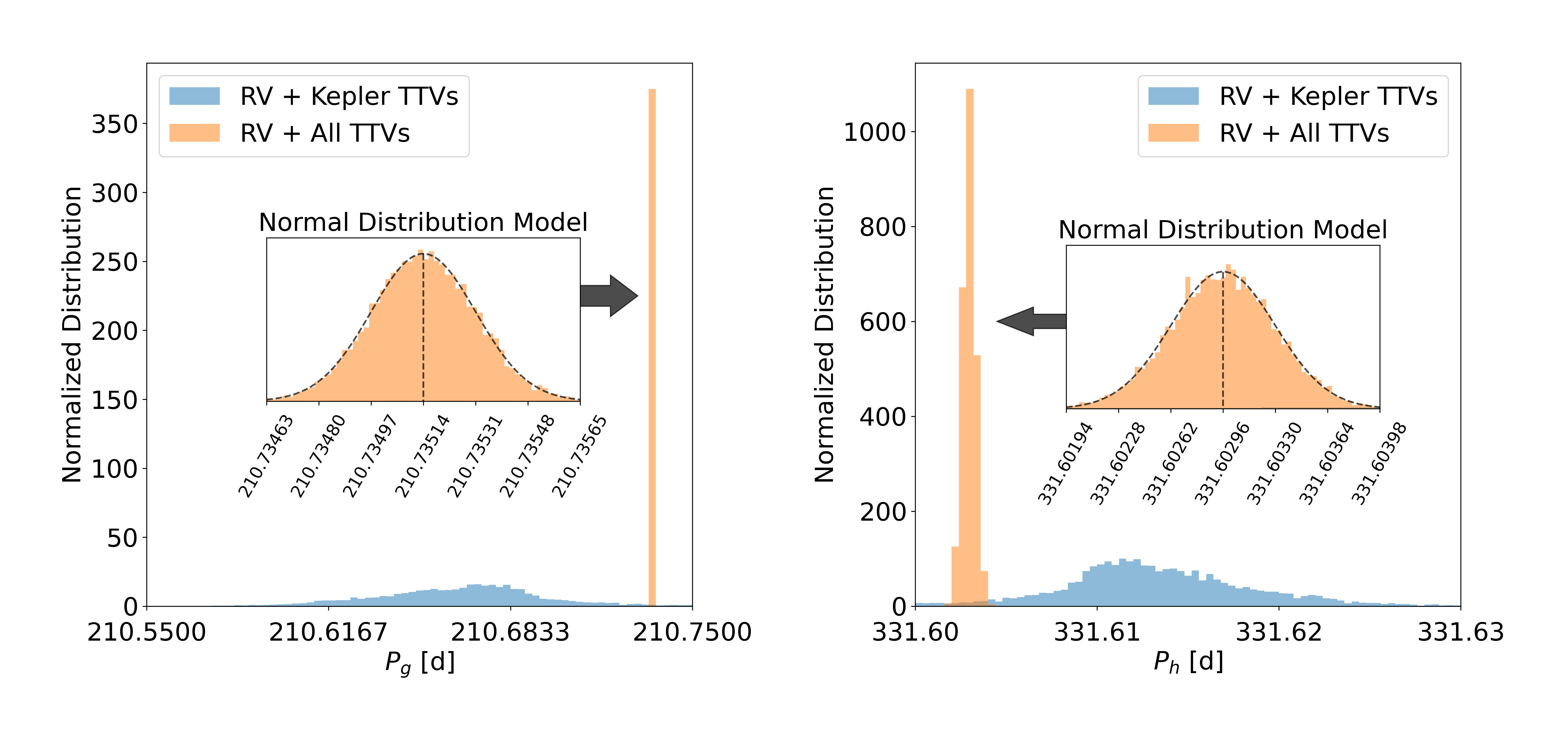}
\caption{Distributions of average orbital periods for K90g (left) and K90h (right) from our joint RV-TTV model including only Kepler transits (blue) and all transits (orange). The histograms are normalized to unit area for consistent comparison. The full dataset's distributions are difficult to visualize when shown on the same scale as those of the Kepler-only dataset, so we present a more narrow binning window in the insets of each plot. We overlay normal distributions with mean and standard deviation equal to the values reported in Table \ref{table:joint-results-table} (which represent the 16th, 50th, and 84th percentiles of the distributions) to show the Gaussian nature of our MCMC posteriors derived from the full dataset.}
\label{fig:period-histograms-plot}
\end{figure*}

The inclusion of post-Kepler transit data improves the estimate precision for planetary orbit parameters: P, T0, e and $\omega$ all have smaller 1$\sigma$ uncertainty intervals in the MCMC analysis of the full RV-TTV dataset compared to those resulting from the Monte Carlo analysis of the RV-Kepler TTV dataset (Table \ref{table:joint-results-table}). The linear ephemeris of K90g in particular improved by several orders of magnitude, owing to the addition of three transit observations and the long time interval between the Kepler-era data and the most recent transit in 2024. The striking difference in period distributions suggested by our model when using the two different datasets is shown in Fig. \ref{fig:period-histograms-plot}. The masses of planets g and h from the updated analysis are each about 1.5$\sigma$ below the median value suggested by the initial analysis of the RVs and Kepler transits. The $\pm 1 \sigma$ uncertainties in the mass of K90g improved slightly whereas those of K90h increased from 4\% to 8\%. Note that our uncertainty in the stellar mass was 8\%, which is consistent with the increased mass uncertainty for planet h. Another possibility is that the increased uncertainty in the mass of planet h is driven by the K90g transit observed with Spitzer, which is a significant outlier. In any case, our updated masses for K90g and K90h are in excellent agreement with those found by \cite{Liang_2021} of $M_g \, = \, 15.0^{+0.9}_{-0.8} \, M_\oplus$ and $M_h \, = \, 203^{+5}_{-5} \, M_\oplus$, although our uncertainty in the mass of planet h ($\sigma_{M_h} \, = \, 16 \, M_\oplus$) is about three times greater than theirs. The reason for the error discrepancy remains unclear. The analyses used inherently different datasets (ours had RVs and more transits; that of \cite{Liang_2021} included transit duration variations), we allowed the stellar mass to vary (they did not), and it is possible that one (or both) samplers under-explored the posterior, resulting in under-reported errors.

As a consistency check, we optimized the N-body model TTVFast \cite{TTVFast} to simultaneously reproduce our observed transits and RVs. We assumed coplanar orbits (inclinations of 90$^\circ$ and longitudes of ascending nodes of 0$^\circ$) and included RV zero-point offset and stellar jitter parameters for the RV portion of the analysis. In doing so, we obtained orbital parameters within 1$\sigma$ of those reported in \cite{Liang_2021} and masses consistent with our results above.  However, note that the N-body parameters correspond to instantaneous orbital elements at the start of an N-body integration, which can differ from their average values.  For the rest of this paper, we refer to the results from our linear ephemeris model (TTVFaster and Radvel, applied to transits and RVs).

\begin{figure*}[ht!]
\plotone{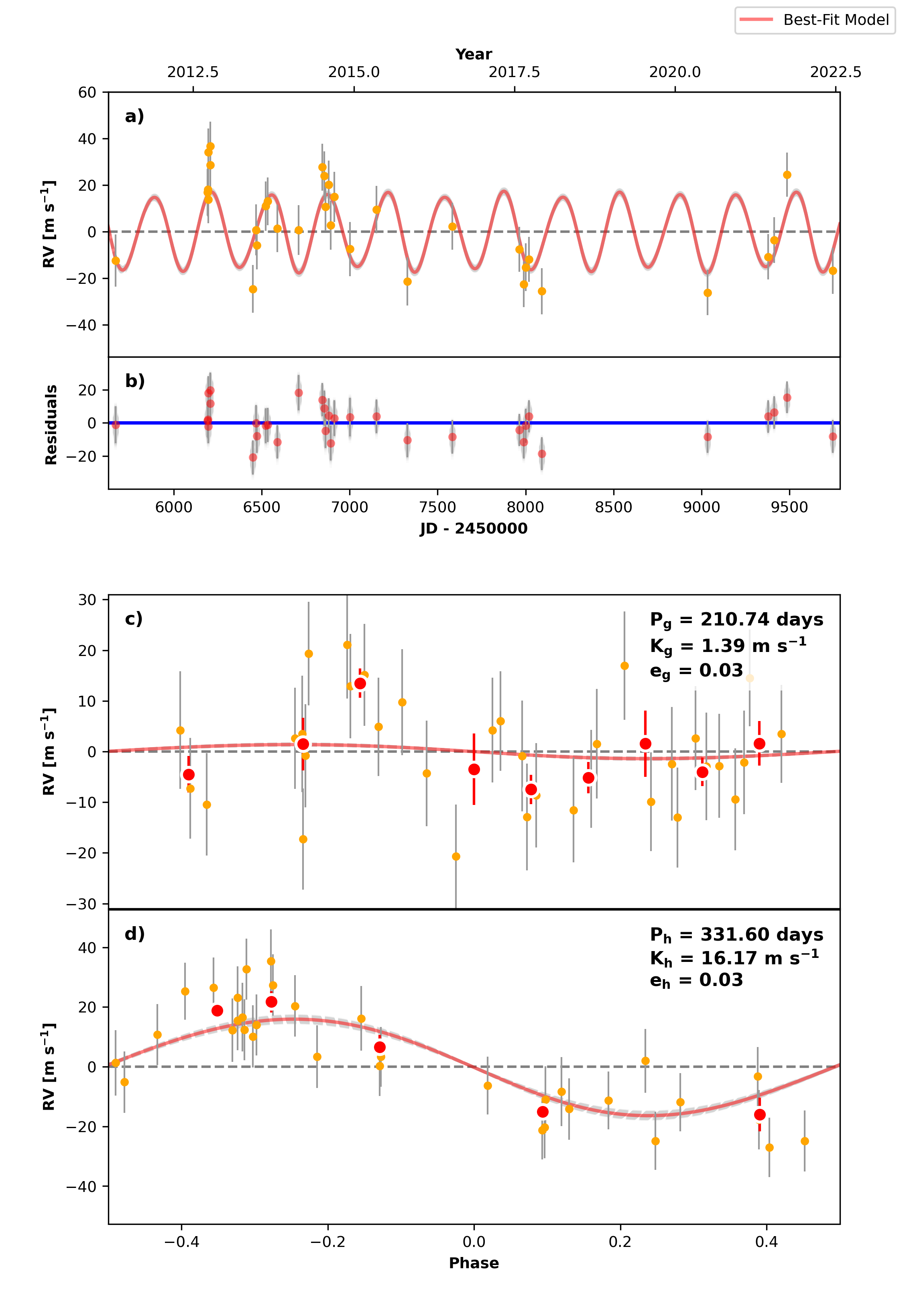}
\caption{Observed RVs for Kepler-90 as in Fig. \ref{fig:rv-data-and-model-spread-joint-analysis-plot} but with our joint model fit to all RVs and TTVs including post-Kepler observations. The solid red line shows the best-fit solution, and the gray dashed lines correspond to 1000 draws from the MCMC posteriors.}
\label{fig:rv-data-fits-mcmc-Mstar}
\end{figure*}

\begin{figure*}[ht!]
\plotone{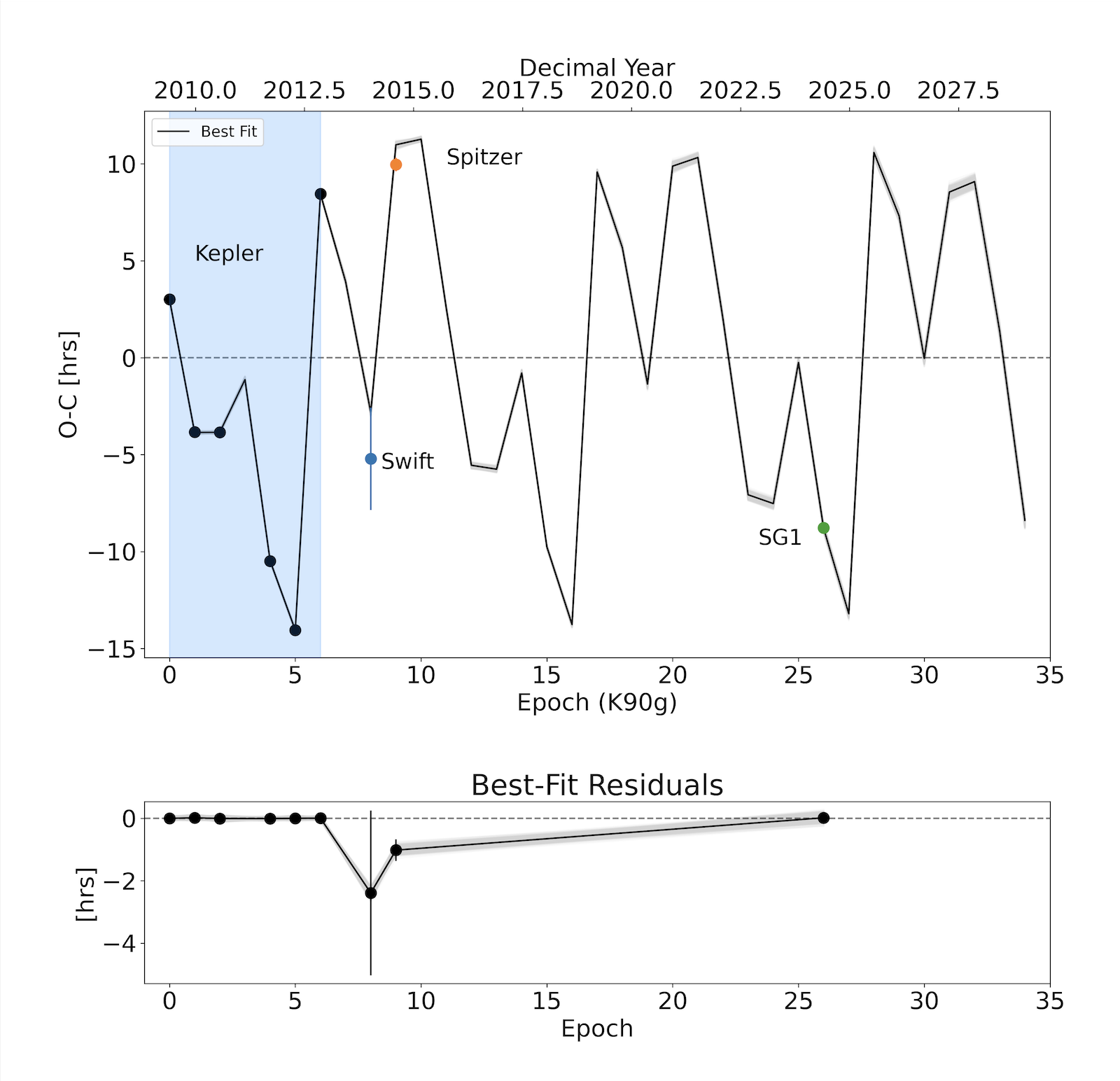}
\caption{Predicted TTVs for 35 epochs of K90g as in Fig. \ref{fig:ttv-predictions-plots-kepler} but with our joint model fit to all RVs and TTVs including post-Kepler observations. The solid line shows the best-fit solution, and the gray lines correspond to 1000 draws from the MCMC posteriors. These predictions extend through January 2029.}
\label{fig:ttv-predictions-plots-all-transits-g}
\end{figure*}

\begin{figure*}[ht!]
\plotone{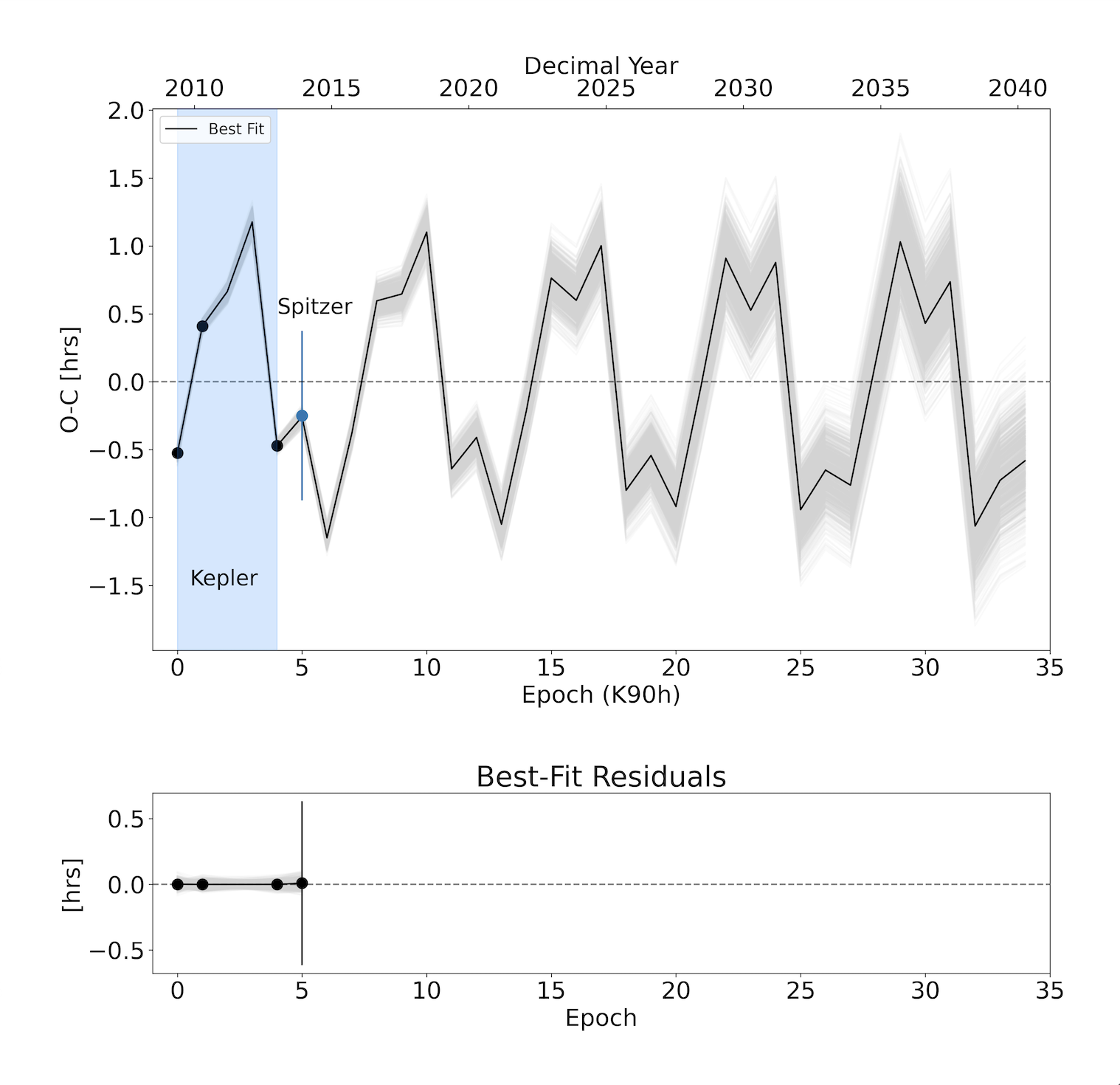}
\caption{Same as Fig. \ref{fig:ttv-predictions-plots-all-transits-g} but for planet h.  The predictions extend through April 2040.}
\label{fig:ttv-predictions-plots-all-transits-h}
\end{figure*}

Our model's best fit to the RVs and a sample of the posterior distribution is shown in Fig. \ref{fig:rv-data-fits-mcmc-Mstar}. The RV data has a clear signal for K90h, but K90g's semi-amplitude is too small to be strongly detected (See \S\ref{subsec:rv-discussion} for more discussion on this point). The best fit to the available TTV data (Figs. \ref{fig:ttv-predictions-plots-all-transits-g} and \ref{fig:ttv-predictions-plots-all-transits-h}) is tightly constrained by the Kepler and ground-based transits ($\sigma_{Kepler} \sim 0.001 \, \textrm{d} \, \textrm{,} \, \sigma_{SG1} \sim 0.003 \, \textrm{d}$), but the uncertainties in the Swift ($\sigma_{Swift} \sim 0.1 \, \textrm{d}$) and Spitzer ($\sigma_{Spitzer} \sim 0.01 \, \textrm{d}$) measurements are 1-2 orders of magnitude larger, and the model's fit to these points is correspondingly worse. The comparatively large range of modeled transit times at these epochs is noticeable in the structure and spread of the residuals in Figs. \ref{fig:ttv-predictions-plots-all-transits-g} and \ref{fig:ttv-predictions-plots-all-transits-h}. While a careful reanalysis of the Swift and Spitzer photometry might reduce the errors, and a full photodynamical analysis working at the level of the photometry itself may lead to a better fit to the data, the most effective way to improve our understanding of the Kepler-90 gas giants' dynamics would be to observe more transits. The long time baseline leading up to the 2024 ground-based observation and its relatively low error significantly reduced the previous model errors by giving a more accurate value for the average period of K90g. After including post-Kepler transit observations, our models suggest that the average period of K90g is $\sim$ 0.03\% ($>$ 1 hour) longer than what was determined from the RV and Kepler data alone. This change in the linear ephemeris along with the large TTVs observed in the system further illustrate the importance of obtaining additional transit observations to better constrain planet properties and future times of transit.  


\section{Discussion} \label{sec:discussion}

\subsection{Relative Contributions of the RV Dataset} \label{subsec:rv-discussion}

\begin{figure*}[ht!]
\plotone{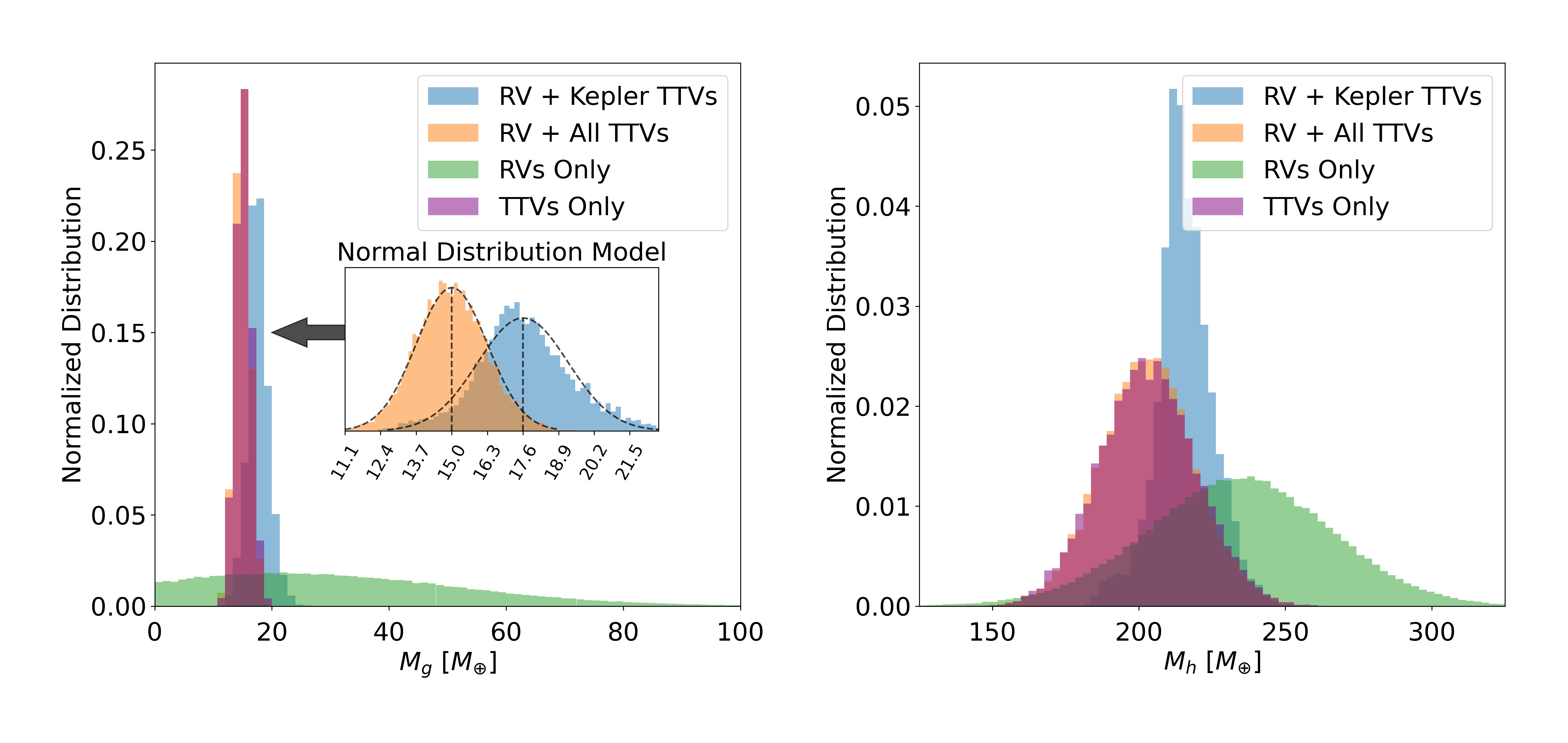}
\caption{Distributions of masses for K90g (left) and K90h (right) from four different models: the joint RV-TTV dataset with only Kepler transits (blue), the joint RV-TTV dataset with all transits (orange), the TTV-only dataset, including post-Kepler transits (purple), and the RV-only dataset (green). The posterior mass distributions of the full RV-TTV analysis are nearly identical to those of the TTV-only analysis. The histograms are all normalized to unit area for consistent comparison. As in Fig. \ref{fig:period-histograms-plot}, normal distributions with mean and standard deviations taken from Table \ref{table:joint-results-table} are plotted over joint RV-TTV distributions in the inset for K90g.}
\label{fig:joint-vs-rv-Mg-vs-Mh-plot}
\end{figure*}

We conducted an MCMC analysis using RadVel to estimate the posteriors for an RV-only likelihood function. Using just the RV data, we fixed planets g and h on Keplerian orbits according to our best fit to the full dataset (Table \ref{table:joint-results-table}) but varied the masses of both planets and the period of h along with the RV parameters $\gamma$ and $\sigma_{\mathrm{jitter}}$. We used 8 ensembles of 128 walkers taking up to 1000 steps each (i.e., up to 1,024,000 possible total steps) and discarded first 50 steps of each walker as burn-in.  We deemed the chains well-mixed after 768,000 steps, based on the MCMC autocorrelation factor ($>50$)and Gelman-Rubin statistic ($<1.01$) for each parameter. The resulting mass posterior distributions for K90g and K90h are compared to those of the joint RV-TTV model in Fig. \ref{fig:joint-vs-rv-Mg-vs-Mh-plot}.

Planet g is essentially undetected in the RVs alone: the mass posteriors for K90g have a large variance and a cutoff at zero due to our positive-mass prior, yielding a 99.7\% confidence upper-limit (3$\sigma$) of $M_g < 109\,M_\oplus$.  Thus, information about the mass of planet g comes almost entirely from TTVs. However, there is a significant signal corresponding to planet h, yielding $M_h = 234^{+31 }_{-32 } \, M_\oplus$.  The RV-only analysis favors a larger mass than our joint model of RVs and TTVs, although the two are statistically consistent as seen by the overlap of the two models' histograms.

\subsection{Relative Contributions of the TTV Dataset} \label{subsec:ttv-discussion}
With the inclusion of the post-Kepler transits, there are 13 observed transits of K90g and K90h. In a model where the planets are coplanar and stellar mass is allowed to vary, there are 11 free parameters (4 orbital parameters per planet, 2 planet masses, and 1 stellar mass), so there is now enough data to fit a TTV-only analysis of these two planets.  We optimized our TTVFaster model (while letting the stellar mass vary) and performed an MCMC analysis in the manner of \S\ref{sec:2024-re-analysis}, but considered only the times of transit in the likelihood function (i.e., we excluded the RVs). The resulting posterior distributions of the planet masses are shown in Fig. \ref{fig:joint-vs-rv-Mg-vs-Mh-plot} in purple. These TTV-only posteriors are nearly identical to those of the joint RV-TTV analysis when all transits are included. This is likely because 1) the stellar mass was allowed to vary in both of these models but was fixed in the others, and 2) the long time baseline provided by the May 2024 transit of K90g essentially ``solves'' the orbit.  The inclusion of the post-Kepler transits provides a much stronger constraint on the system than the decade of RV signals, especially with the non-detection of planet g in the RVs. This result further emphasizes the importance of observing additional transits of long-period planets exhibiting TTVs, particularly at times that extend the baseline and constrain the linear ephemeris.  However, it is noteworthy that our ability to recover the May 2024 transit significantly benefited from our initial joint RV+TTV analysis (blue), such that the RVs were critical in enabling the eventual characterization of the full set of TTVs (purple and orange).

\subsection{Sensitivity to Companions Beyond 1 AU}
\label{subsec:rv-companions-discussion}

\begin{figure*}[ht!]
\plotone{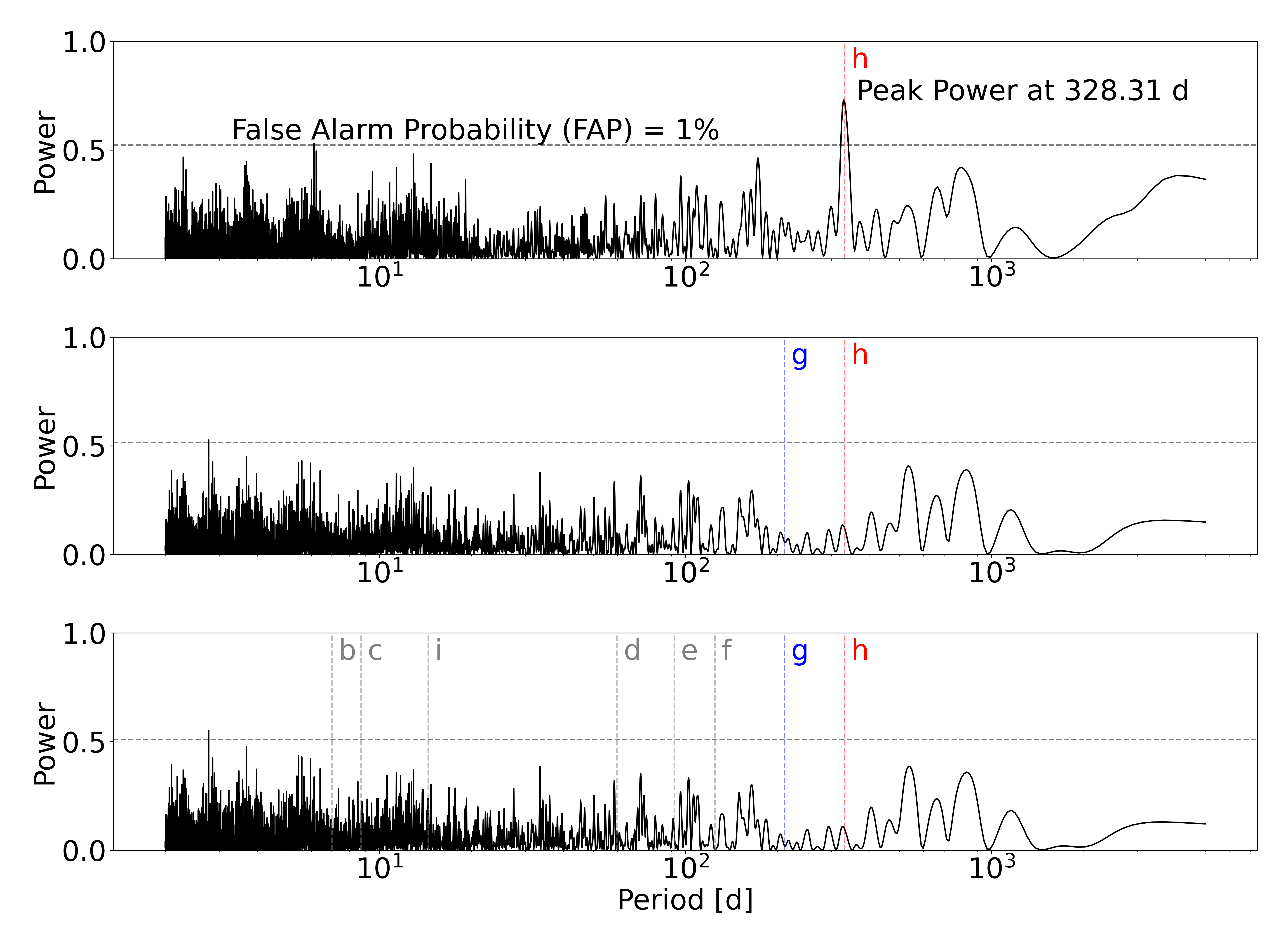}
\caption{(Top) Lomb-Scargle (LS) periodogram of the 34 RV measurements over an orbital period range of 2-5000 days. There is a clear peak in power near the orbital period of planet h, while no other signal is detected above a 1\% false-alarm probability. (Middle) LS periodogram of residual RV data after subtraction of planet h's RV model based on the best-fit parameters. Even after subtraction, no clear signal is seen at the period of planet g. (Bottom) LS periodogram of residual RV data after subtraction of both planet h and g's RV models based on best joint fit parameters. The power peak at 2.77 d is not from any known planet and is likely from noise and/or aliasing.}
\label{fig:periodogram-plot}
\end{figure*}

After modeling the RV signals of planets g and h with a Keplerian orbit, the Lomb-Scargle Periodogram (\citealt{lomb1976}, \citealt{scargle1982}; implemented via Astropy \citealt{astropy2022}) of the residuals does not contain any peaks above the $1\%$ false-alarm probability (determined via the bootstrap method) beyond the orbit of K90h (Fig. \ref{fig:periodogram-plot}). Because the probability of observing a transit decreases with the semimajor axis of a planet's orbit purely from geometry \citep{winn2010}, long-period planets are unlikely to be detected in transit photometry. \cite{Weiss2024} used the non-detection of a linear trend in the RVs to place an upper bound on additional companions of $M\mathrm{sin}i < 1.8\,M_J$ at 10 AU ($< 0.5\,M_J$ at 5 AU) at 3$\sigma$ confidence.


\subsection{Comparison with the Solar System} \label{subsec:sol-comparison}

\begin{figure*}[ht!]
\plotone{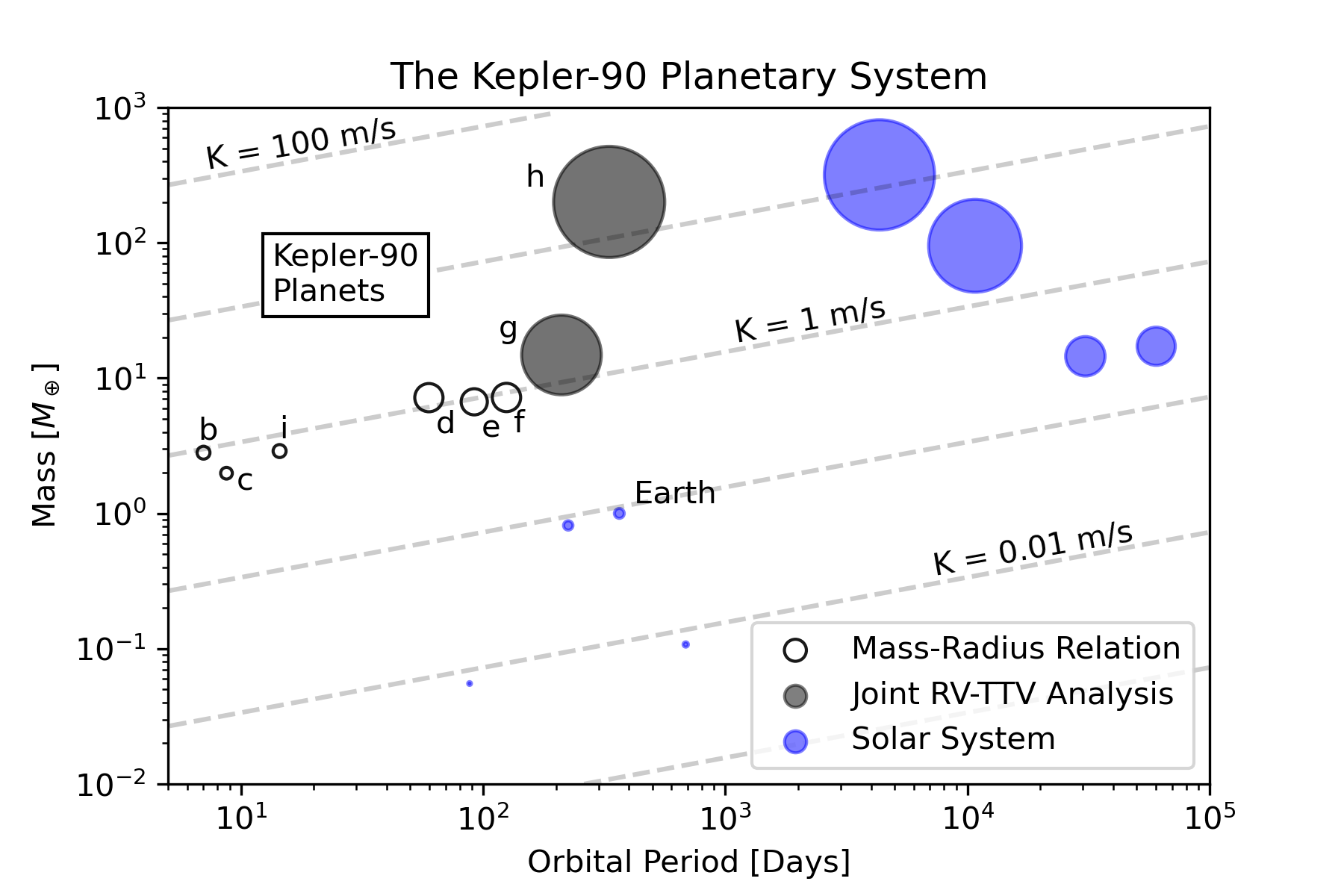}
\caption{Comparison of the Kepler-90 system (gray) with our own solar system (blue). The size of each circle is proportional to the planet's radius. Masses for the inner six Kepler-90 planets are estimated using their measured radii and an empirical mass-radius relation \citep{mass-radius-relation}. Error bars are generally too small to be seen on this scale. Lines of constant RV semi-amplitude are drawn for a solar-mass star and circular orbit to approximately illustrate the  RV signal induced by each planet.}
\label{fig:k90-sol-comparison-plot}
\end{figure*}

In some ways the Kepler-90 system arguably bears the closest resemblance to our own solar system of any host star found to date. K90 is a sun-like star with eight planets of mixed size having a total mass within about a factor of two of the solar system. Despite this comparable total mass, all of the planetary orbits in the K90 system lie within 1 AU of the host star (Fig. \ref{fig:k90-sol-comparison-plot}). Additionally, both systems have an adjacent pair of Saturn and Jupiter sized planets exterior to a set of smaller inner planets.

A key difference of the K90 system is the ordering of the pair of ``giant'' planets ($R > 6\,R_\oplus$). In the K90 system they are inverted relative to the solar system: the lower-mass giant planet has the shorter period. Additionally the Saturnian analogue (K90g) has about 25\% the density of Saturn. This inversion was almost certainly significant in the architectural evolution of K90 compared to the solar system. According to one model, Jupiter and Saturn's crossing of resonances during their inward migrations excited the giant planets and led to their current orbital architecture \citep{tsiganis2005}. In contrast, there are no (known) giant planets beyond the Jupiter-analogue K90h.

Additionally, an inward-then-outward migration of Jupiter and Saturn called the Grand Tack may have been responsible for disrupting the inner planetesimal disk during the early evolution of the solar system, possibly destroying any first-generation planets and allowing for the formation of the ``second generation" observed rocky worlds near/in the habitable zone that we see today \citep{walsh2011, batygin2015}. The Grand Tack mechanism would not have been triggered in the inverted architecture of the K90 giant planets, and so it is possible that the K90 inner planets are primordial.  The six planets interior to K90g and K90h have a diversity of sizes with radii $1.2 \, R_\oplus < R < 3 \, R_\oplus$ and likely masses $1 \, M_\oplus < M < 10 \, M_\oplus$. In particular the planets larger than $2 R_\oplus$ almost certainly have volatile envelopes, a feature absent from the inner solar system. A thorough dynamical investigation is needed to better understand what effects, if any, the inverted mass ordering of the K90 giant planets and the low-density of K90g produced among K90's inner planets.

\subsection{Future Opportunities for Observations} \label{subsec:future-obs}

Given the exceptional nature of the Kepler-90 system and its similarity to our own solar system, we would very much like to probe the chemical abundances in the system via e.g. transmission spectroscopy of planetary atmospheres. Kepler-90 g and h's large transit depths suggest that it should be easy to detect the planets' atmospheric absorption, making them an ideal pair of cool Jupiter-Saturns to probe with the James Webb Space Telescope and Hubble Space Telescope in the manner of \cite{JWSTspec}, \cite{libby-roberts2020}, and \cite{HIP41378}. We have demonstrated that the RV and Kepler data alone do not constrain future times of transit precisely enough to predict a transit midpoint with $1\sigma$ uncertainty less than the transit duration itself (12.6 hours and 14.7 hours for K90g and K90h respectively). However, with the inclusion of the post-Kepler transit observations presented above, the $1\sigma$ uncertainties on transit midpoints through 2029 are $\sim10$ minutes for both K90g and K90h. These transit midpoint predictions are reported in Table \ref{table:predicted-transits} through 35 epochs for each planet. It should be noted that these predicted times of transit were made using TTVFaster (rather than an N-body model) and by modeling only the interactions between K90g and K90h. The perturbing effects of other planets in the system (either known or unknown) may slightly inflate the uncertainties on our values. Nevertheless, this dramatic improvement in timing precision makes planning future observations of the Kepler-90 giant planets a feasible endeavor.

\begin{deluxetable*}{ccccc}
\label{table:predicted-transits}
\tablecaption{35 Transit Midpoint Predictions for K90g and K90h}

\tablehead{ \colhead{Epoch} & \colhead{K90g Predicted Transits} & \colhead{K90g O-C} & \colhead{K90h Predicted Transits} & \colhead{K90h O-C} \\ 
\colhead{} & \colhead{[BJD - 2454900]} & \colhead{[Minutes]} & \colhead{[BJD - 2454900]} & \colhead{[Minutes]} } 

\startdata
     0 &                        $80.101\pm0.001$ &       $181\pm2$ &                        $73.479\pm0.001$ &         $-31\pm2$ \\
     1 &                        $290.55\pm0.002$ &      $-232\pm2$ &                       $405.121\pm0.001$ &          $25\pm1$ \\
     2 &                       $501.286\pm0.002$ &      $-231\pm3$ &                       $736.734\pm0.001$ &          $40\pm2$ \\
     3 &                       $712.135\pm0.003$ &       $-68\pm4$ &                      $1068.358\pm0.002$ &          $70\pm3$ \\
     4 &                        $922.48\pm0.001$ &      $-629\pm2$ &                      $1399.893\pm0.001$ &         $-28\pm1$ \\
     5 &                      $1133.066\pm0.001$ &      $-843\pm2$ &                      $1731.505\pm0.001$ &         $-15\pm2$ \\
     6 &                      $1344.739\pm0.001$ &       $506\pm2$ &                      $2063.071\pm0.002$ &         $-69\pm3$ \\
     7 &                      $1555.286\pm0.003$ &       $236\pm4$ &                      $2394.706\pm0.002$ &         $-23\pm3$ \\
     8 &                       $1765.74\pm0.004$ &      $-169\pm6$ &                      $2726.349\pm0.003$ &          $36\pm4$ \\
     9 &                      $1977.049\pm0.003$ & $658^{+4}_{-5}$ &                      $3057.954\pm0.003$ &          $39\pm4$ \\
    10 &                      $2187.797\pm0.002$ &       $676\pm3$ &                      $3389.576\pm0.004$ &          $66\pm5$ \\
    11 &                       $2398.17\pm0.002$ &       $154\pm3$ &                      $3721.106\pm0.003$ &         $-38\pm4$ \\
    12 &                      $2608.566\pm0.003$ &      $-333\pm4$ &                      $4052.719\pm0.003$ &         $-24\pm5$ \\
    13 &                      $2819.293\pm0.003$ &      $-345\pm4$ &                      $4384.295\pm0.004$ &         $-63\pm5$ \\
    14 &                      $3030.235\pm0.003$ &       $-47\pm4$ &                      $4715.933\pm0.004$ &         $-13\pm6$ \\
    15 &                      $3240.597\pm0.002$ &      $-585\pm3$ &                      $5047.577\pm0.005$ &          $46\pm7$ \\
    16 &                      $3451.165\pm0.002$ &      $-825\pm3$ &                      $5379.173\pm0.005$ &    $36^{+7}_{-8}$ \\
    17 &                      $3662.872\pm0.002$ &       $574\pm3$ &                      $5710.793\pm0.006$ &          $60\pm8$ \\
    18 &            $3873.446^{+0.004}_{-0.003}$ &       $341\pm5$ &                      $6042.321\pm0.005$ &         $-48\pm8$ \\
    19 &            $4083.887^{+0.005}_{-0.004}$ & $-81^{+7}_{-6}$ &                      $6373.934\pm0.006$ &         $-32\pm8$ \\
    20 &                       $4295.09\pm0.004$ &       $592\pm6$ &                      $6705.522\pm0.006$ &         $-55\pm9$ \\
    21 &                      $4505.844\pm0.004$ &       $619\pm6$ &                      $7037.161\pm0.007$ &   $-2^{+9}_{-10}$ \\
    22 &                      $4716.232\pm0.004$ &       $120\pm6$ &            $7368.804^{+0.007}_{-0.008}$ &         $55\pm11$ \\
    23 &                      $4926.589\pm0.004$ &      $-425\pm6$ &            $7700.391^{+0.007}_{-0.008}$ &         $32\pm11$ \\
    24 &                      $5137.306\pm0.004$ &      $-451\pm6$ &                      $8032.008\pm0.008$ &         $53\pm12$ \\
    25 &                      $5348.344\pm0.004$ &       $-15\pm6$ &                      $8363.535\pm0.008$ &        $-56\pm11$ \\
    26 &                      $5558.723\pm0.003$ &      $-528\pm5$ &                       $8695.15\pm0.008$ &        $-39\pm12$ \\
    27 &                      $5769.275\pm0.004$ &      $-792\pm6$ &                      $9026.749\pm0.008$ &        $-45\pm12$ \\
    28 &                      $5981.001\pm0.004$ &       $634\pm6$ &                      $9358.389\pm0.009$ &          $8\pm13$ \\
    29 &                        $6191.6\pm0.005$ &       $439\pm7$ &                       $9690.029\pm0.01$ &         $62\pm14$ \\
    30 &                      $6402.029\pm0.006$ &        $-2\pm8$ &                      $10021.607\pm0.01$ &         $26\pm14$ \\
    31 &                      $6613.121\pm0.006$ &       $511\pm9$ &                      $10353.223\pm0.01$ &         $44\pm15$ \\
    32 &                      $6823.879\pm0.006$ &       $545\pm8$ &                      $10684.751\pm0.01$ &        $-63\pm14$ \\
    33 &                      $7034.292\pm0.006$ &        $81\pm8$ &                      $11016.368\pm0.01$ &        $-43\pm15$ \\
    34 &                      $7244.621\pm0.006$ &      $-504\pm8$ &                     $11347.977\pm0.011$ & $-35^{+15}_{-16}$ \\
\enddata

\tablecomments{All entries are the Median and $\pm 1\sigma$ values from the marginalized MCMC posteriors corresponding to the full RV-TTV dataset described in \S\ref{sec:2024-re-analysis}. O-C values were computed by subtracting the best-fit linear ephemeris from TTVFaster's modeled transit times. The final epoch reported occurs in January 2029 for K90g and April 2040 for K90h.}

\end{deluxetable*}

\section*{Acknowledgments}
\label{sec:acknowledgments}
We thank Jason Rowe, Brett Morris, and Shreyas Vissapragada for helpful discussions.  We thank the astronomers who attempted to observe the May 2024 transit of Kepler-90g but were unsuccessful, including Peter Garnavich, Jonathan Crass, and Valeria Bautista. We thank Kevin B. Stevenson and Hannah Diamond-Lowe for their contributions to processing the Spitzer transits and Brett Morris for processing the Swift transit used in this work.

L.M.W.\ acknowledges support from the NASA Exoplanet Research Program through grant 80NSSC23K0269 and from NASA-Keck Key Strategic Mission Support Grant No. 80NSSC19K1475.  E.A.\ acknowledges support from NSF grant No. AST-1907342, NASA NExSS grant No. 80NSSC18K0829, and NASA XRP grant No. 80NSSC21K1111.
The postdoctoral fellowship of KB is funded by F.R.S.-FNRS grant T.0109.20 and by the Francqui Foundation.
MG is F.R.S.-FNRS Research Director.
Adam Popowicz was financed by grant 02/140/RGJ24/0031 from the Silesian University of Technology.
This work is partly supported by JSPS KAKENHI Grant Number JP24H00017 and JSPS Bilateral Program Number JPJSBP120249910.

This work is based in part on observations made with the Spitzer Space Telescope, which was operated by the Jet Propulsion Laboratory, California Institute of Technology under a contract with NASA. Support for this work was provided by NASA through an award issued by JPL/Caltech.

This paper is based on observations made with the MuSCAT instruments, developed by the Astrobiology Center (ABC) in Japan, the University of Tokyo, and Las Cumbres Observatory (LCOGT). MuSCAT3 was developed with financial support by JSPS KAKENHI (JP18H05439) and JST PRESTO (JPMJPR1775), and is located at the Faulkes Telescope North on Maui, HI (USA), operated by LCOGT. MuSCAT4 was developed with financial support provided by the Heising-Simons Foundation (grant 2022-3611), JST grant number JPMJCR1761, and the ABC in Japan, and is located at the Faulkes Telescope South at Siding Spring Observatory (Australia), operated by LCOGT.

Data were also obtained in part using the 1.3\,m McGraw-Hill Telescope of the MDM Observatory, with observing assistance from J. Escalera, G. Garcia, and C. Pumpo. MDM is operated by Dartmouth College, Columbia University, The Ohio State University, Ohio University, and the University of Michigan.

This research has made use of the NASA Exoplanet Archive, which is operated by the California Institute of Technology, under contract with the National Aeronautics and Space Administration under the Exoplanet Exploration Program.

\facilities{Kepler, Keck:I (HIRES), LCOGT, FTN (MuSCAT3), Swift, Spitzer}

\software{Astropy \citep{astropy2022}, TTVFaster \citep{ttvFaster}, TTVFast \citep{TTVFast}, RadVel \citep{radvel}, SciPy \citep{2020SciPy-NMeth}, emcee3 \citep{emcee}, AstroImageJ \citep{Collins:2017}, {\tt TRAFIT} \citep{Gillon2010AA,Gillon2012,Gillon2014AA}}

\bibliography{k90-bib}{}
\bibliographystyle{aasjournal}

\end{document}